\documentclass[aps,prb,twocolumn,floatfix,showpacs,10pt,letterpaper]{revtex4}

\pdfoutput=1

\newcommand{\be}{\begin{equation}}
\newcommand{\ee}{\end{equation}}

\begin{document}

\begin{titlepage}

\title{
Symmetry protected topological phases
in non-interacting fermion systems
}

\author{Xiao-Gang Wen}
\affiliation{Department of Physics, Massachusetts Institute of
Technology, Cambridge, Massachusetts 02139, USA}

\begin{abstract}
Symmetry protected topological (SPT) phases are gapped quantum phases with a
certain symmetry, which can all be smoothly connected to the same trivial
product state if we break the symmetry.  For non-interacting fermion systems
with time reversal ($\hat T$), charge conjugation ($\hat C$), and/or $U(1)$
($\hat N$) symmetries, the total symmetry group can depend on the relations
between those symmetry operations, such as $\hat T \hat N \hat T^{-1}= \hat
N$ or $\hat T \hat N \hat T^{-1}=- \hat N$.  As a result, the SPT phases of
those fermion systems with different symmetry groups have different
classifications.  In this paper, we use Kitaev's K-theory approach to classify
the gapped free fermion phases for those possible symmetry groups.  In
particular, we can view the $U(1)$ as a spin rotation.
We find that superconductors with the $S_z$ spin rotation symmetry are
classified by $\Z$ in even dimensions, while superconductors with the time
reversal plus the $S_z$ spin rotation symmetries are classified by $\Z$ in odd
dimensions.  We show that all 10 classes of gapped free fermion phases can be
realized by electron systems with certain symmetries.  We also point out that
to properly describe the symmetry of a fermionic system, we need to specify its
full symmetry group that includes the fermion number parity transformation
$(-)^{\hat N}$. The full symmetry group is actually a projective symmetry group.

\end{abstract}


\maketitle

\end{titlepage}

{\small \setcounter{tocdepth}{1} \tableofcontents }

\begin{table*}[tb]
 \centering
 \begin{tabular}{ |c|c| }
 \hline
Electron systems & Full symm. group $G_f$ \\
 \hline
 \hline
\tbox{5.5in}{
Insulators with spin-orbital coupling
and spin order (or non-coplanar spin order)\\
$ 
(\imth \hat c_i^\dag \v n_1 \cdot \v \si \hat c_{i'}
+\imth \hat c_j^\dag \v n_2 \cdot \v \si \hat c_{j'}
+\imth \hat c_k^\dag \v n_3 \cdot \v \si \hat c_{k'})
+ (\hat c_i^\dag \v n_1 \cdot \v \si \hat c_i
+ \hat c_j^\dag \v n_2 \cdot \v \si \hat c_j
+ \hat c_k^\dag \v n_3 \cdot \v \si \hat c_k)
$
}
& \blue{$U(1)$} \\
\hline
\tbox{5.5in}{
Superconductors  with spin-orbital coupling
and spin order (or non-coplanar spin order)\\
$ 
 \hat c_i^\dag \v n_1 \cdot \v \si \hat c_i
+ \hat c_j^\dag \v n_2 \cdot \v \si \hat c_j
+ \hat c_k^\dag \v n_3 \cdot \v \si \hat c_k
+(\hat c_{\up i}\hat c_{\down j}- \hat c_{\down i}\hat c_{\up j})
$
}
& \blue{``none''$=Z_2^f$} \\
\hline
\tbox{5.5in}{
Insulators with spin-orbital coupling 
$ 
\imth \hat c_i^\dag \v n_1 \cdot \v \si \hat c_{i'}
+\imth \hat c_j^\dag \v n_2 \cdot \v \si \hat c_{j'}
+\imth \hat c_k^\dag \v n_3 \cdot \v \si \hat c_{k'}
$\\
(\emph{symmetry}: charge conservation and time reversal symmetries) 
}
& $G_-^-(U,T)$ \\
 \hline
\tbox{5.5in}{
Superconductors with spin-orbital coupling and real pairing\\
$\imth \hat c_i^\dag \v n_1 \cdot \v \si \hat c_{i'}
+\imth \hat c_j^\dag \v n_2 \cdot \v \si \hat c_{j'}
+\imth \hat c_k^\dag \v n_3 \cdot \v \si \hat c_{k'}
+(\hat c_{\up i}\hat c_{\down j}- \hat c_{\down i}\hat c_{\up j})$
(\emph{symmetry}: time reversal symmetry)
}
& \blue{$G_-(T)=Z_4$} \\
 \hline
\tbox{5.5in}{
Superconductors with $S_z$ conserving spin-orbital coupling and real pairing\\
$\imth \hat c_i^\dag  \si^z \hat c_{j}
+(\hat c_{\up i}\hat c_{\down j}- \hat c_{\down i}\hat c_{\up j})$
(\emph{symmetry}: time reversal and $S_z$ spin rotation symmetries)
}
& \blue{$G_-^+(U,T)=U(1)\times Z^T_2$} \\
\hline
\tbox{5.5in}{
Superconductors with coplanar spin order and real pairing
$ \hat c_i^\dag \v n_1 \cdot \v \si \hat c_i
+\hat c_j^\dag \v n_2 \cdot \v \si \hat c_j
+(\hat c_{\up i}\hat c_{\down j}- \hat c_{\down i}\hat c_{\up j})$\\
(\emph{symmetry}: a combined time reversal and $180^\circ$ spin rotation symmetry) }
 & \blue{$G_+(T)=Z^T_2\times Z_2^f$} \\
\hline
\tbox{5.5in}{
Superconductors with real pairing and collinear spin order
$\hat c_i^\dag\si^z \hat c_j
+(\hat c_{\up i}\hat c_{\down j}- \hat c_{\down i}\hat c_{\up j})$\\
(\emph{symmetry}: $S_z$ spin rotation and a combined time reversal and $180^\circ$ $S_y$ spin rotation symmetry) }
 & $G_+^-(U,T)=U(1)\rtimes Z^T_2$ \\
 \hline
\tbox{5.5in}{Insulators with coplanar spin order
$ \hat c_i^\dag \v n_1 \cdot \v \si \hat c_i
+\hat c_j^\dag \v n_2 \cdot \v \si \hat c_j $\\
(\emph{symmetry}: charge conservation and a combined time reversal and $180^\circ$ spin rotation 
symmetries) }
 & $G_+^-(U,T)=U(1)\rtimes Z^T_2$ \\
\hline
\tbox{5.5in}{Superconductors with real triplet $S_z=0$ paring 
$\hat c_{\up i}\hat c_{\down j}+ \hat c_{\down i}\hat c_{\up j}$\\
(\emph{symmetry}: a combined $180^\circ$ $S_y$ spin rotation and time reversal  symmetry, 
a combined $180^\circ$ $S_y$ spin rotation and charge rotation symmetry, 
and $S_z$-spin rotation symmetry)
} 
& 
$G_{++}^{--}(U,T,C)$\\
 \hline
\tbox{5.5in}{
Superconductors with time reversal,\\
$180^\circ$ $S_y$-spin rotation, and $S_z$-spin rotation symmetries} 
& $G_{--}^{++}(U,T,C)$ \\
\hline
\tbox{5.5in}{
Superconductors with real singlet pairing 
$\hat c_{\up i}\hat c_{\down j}- \hat c_{\down i}\hat c_{\up j}$\\
(\emph{symmetry}: time reversal and $SU(2)$ spin rotation symmetries) }
& $G[SU(2),T]$   \\
\hline
Superconductors with 
$180^\circ$ $S_y$-spin rotation
and $S_z$-spin rotation symmetries
& $G_{-}(U,C)$ \\
\hline
\tbox{5.5in}{
Superconductors with complex singlet pairing
$\e^{\imth \th_{ij}} (\hat c_{\up i}\hat c_{\down j}- \hat c_{\down i}\hat c_{\up j})$\\
(\emph{symmetry}: $SU(2)$-spin rotation symmetry)
}
& $SU(2)$ \\
 \hline
\tbox{5.5in}{
Insulators with spin-orbital coupling 
and inter-sublattice hopping
$ 
\imth \hat c_{i_A}^\dag \v n_1 \cdot \v \si \hat c_{i_B}
+\imth \hat c_{j_A}^\dag \v n_2 \cdot \v \si \hat c_{j_B}
+\imth \hat c_{k_A}^\dag \v n_3 \cdot \v \si \hat c_{k_B}
$\\
(\emph{symmetry}: charge conservation, time reversal and charge conjugation symmetries) 
}
& $G_{--}^{-+}(U,T,C)$ \\
\hline
 \end{tabular}
\caption{(Color online) Electron systems and their full symmetry groups $G_f$.
The groups are defined in Table \ref{groups}.  The symmetry group symbols have
the following meaning: for example, $G_{--}^{++}(U,T,C)$ is a symmetry group
generated by $\hat N$ (the $U(1)$ fermion number conservation or spin
rotation), $\hat T$ (the time reversal), and $\hat C$ (the charge conjugation
or $180^\circ$ spin rotation).  The $\pm$ subscripts/superscripts describe the
relations between the transformations $\hat N$, $\hat T$, and/or $\hat C$ (see
Table \ref{groups}).  Some times, when we describe the symmetry of a fermion
system, we do not include the fermion number parity transformation $(-)^{\hat
N}$ in the symmetry group $G$.  Here the full symmetry group $G_f$ does include
the fermion number parity transformation $(-)^{\hat N}$.  So the full  symmetry
group of a fermion system with no symmetry is $G_f=Z_2^f$ generated by the
fermion number parity transformation.  $G_f$ is a $Z_2^f$ extension of $G$:
$G=G_f/Z_2^f$.  Free electron systems with symmetry $G_{--}^{++}(U,T,C)$
actually have a higher symmetry $G[SU(2),T]$.  Similarly, free electron systems
with symmetry $G_-(U,C)$ actually have a higher symmetry $SU(2)$.  The groups
in blue are Abelian and electron operators form their 1D representations.
}
 \label{elegrp}
\end{table*}

\section{Introduction}

We used to believe that all possible phases and phase transitions are described
by Landau symmetry breaking theory.\cite{L3726,GL5064,LanL58}  However, the
experimental discovery of fractional quantum Hall states\cite{TSG8259,L8395}
and the theoretical discovery of chiral spin liquids\cite{KL8795,WWZ8913}
indicate that new states of quantum matter without symmetry breaking and
without long range order can exist. Such new kind of orders is called
topological order,\cite{WNtop,Wrig} since their low energy effective theories
are topological quantum field theories.\cite{W9038} At first, the theory of
topological order was developed based on their robust ground state degeneracy
on compact spaces and the associated robust non-Abelian Berry's
phases.\cite{WNtop,Wrig} Later, it was realized that topological order can be
characterized by the boundary excitations,\cite{H8285,Wedgerev} which can be
directly probed by experiments.  One can develop a theory of topological
order based on the boundary theory.\cite{Wtoprev}

\begin{table*}[t]
 \centering
 \begin{tabular}{ |c||c|c|c|c|c|c|c|c|c|c|c| }
 \hline
Symmetry & $C_p|_{\text{for }d=0}$ & $p \backslash d$
  & 0 & 1 & 2 & 3 & 4 & 5 & 6 & 7 &
example
\\
\hline
\hline
{
\footnotesize
$\bmm
\red{U(1)}\\[1mm]
G_-(C)
\emm$
} & $\frac{U(l+m)}{U(l)\times U(m)}\times \Z$ & 0 
& $\Z$ & $0$ & $\Z$ & $0$ & $\Z$ & $0$ & $\Z$ & $0$ &
$\bmm
\text{(Chern)}\\
\text{insulator}
\emm$
~~
$\bmm
\text{supercond.}\\[-1mm]
\text{with collinear}\\[-1mm]
\text{spin order}
\emm$
\\
 \hline
{
\footnotesize
$\bmm
\red{G_\pm^+(U,T)}\\[1mm]
G_{--}^+(T,C)\\[1mm]
G_{+-}^+(T,C)
\emm$
}  & $U(n)$ & 1
& $0$ & $\Z$ & $0$ & $\Z$ & $0$ & $\Z$ & $0$ & $\Z$ &
$\bmm
\text{supercond. w/ real pairing}\\[-1mm]
\text{and $S_z$ conserving}\\[-1mm]
\text{spin-orbital coupling}
\emm$
\\
\hline
 \end{tabular}
\caption{(Color online) Classification of the gapped phases of non-interacting
fermions in $d$-dimensional space, for some symmetries.  The space of the
gapped states is given by $C_{p+d \text{ mod }2}$, where $p$ depends on the
symmetry group.  The distinct phases are given by $\pi_0(C_{p+d\text{ mod
}2})$.  ``0'' means that only trivial phase exist.  $\Z$ means that non-trivial
phases are labeled by non-zero integers and the trivial phase is labeled by 0.
The groups in red can be realized by electron systems (see Table \ref{elegrp}).
}
 \label{GFF}
\end{table*}
\begin{table*}[tb]
{
 \begin{tabular}{ |c||c|c|c|c|c|c|c|c| }
 \hline
Symm.
&
{\scriptsize
$\bmm
\red{G_{+}^{-}(U,T)}\\[1mm]
G_{+-}^{-}(T,C)
\emm$
}%
& 
{\scriptsize
$\bmm
\red{G_+(T)}\\[1mm]
G_{++}^{+}(T,C)\\[1mm]
\red{G_{++}^{--}(U,T,C)}\\[1mm]
G_{++}^{-+}(U,T,C)\\[1mm]
G_{-+}^{+-}(U,T,C)\\[1mm]
G_{++}^{++}(U,T,C)
\emm$
}%
&
{\scriptsize
$\bmm
\text{{\small\red{``none''}}}\\[1mm]
G_+(C)\\[1mm]
G_{++}^{-}(T,C)\\[1mm]
G_{-+}^{-}(T,C)\\[1mm]
G_+(U,C)
\emm$
}%
&
{\scriptsize
$\bmm
\red{G_-(T)}\\[1mm]
G_{-+}^{+}(T,C)\\[1mm]
G_{-+}^{--}(U,T,C)\\[1mm]
G_{-+}^{-+}(U,T,C)\\[1mm]
G_{++}^{+-}(U,T,C)\\[1mm]
G_{-+}^{++}(U,T,C)
\emm$
}%
&
{\scriptsize
$\bmm
\red{G_{-}^{-}(U,T)}\\[1mm]
G_{--}^{-}(T,C)
\emm$
}%
&
{\scriptsize
$\bmm
G_{--}^{--}(U,T,C)\\[1mm]
\red{G_{--}^{-+}(U,T,C)}\\[1mm]
G_{--}^{+-}(U,T,C)\\[1mm]
G_{+-}^{++}(U,T,C)
\emm$
}%
&
{\scriptsize
$\bmm
\red{G_{-}(U,C)}\\[1mm]
\red{SU(2)}
\emm$
}
&
{\scriptsize
$\bmm
G_{+-}^{--}(U,T,C)\\[1mm]
G_{+-}^{-+}(U,T,C)\\[1mm]
G_{+-}^{+-}(U,T,C)\\[1mm]
\red{G_{--}^{++}(U,T,C)}\\[1mm]
\red{G[SU(2),T]}
\emm$
}%
\\
 \hline
$R_p|_{\text{for }d=0}$ &
{\footnotesize
$\frac{O(l+m)}{O(l)\times O(m)}\times \Z$
}
& 
 $O(n)$ & 
 $\frac{O(2n)}{U(n)}$ & 
 $\frac{U(2n)}{Sp(n)}$ &
{\footnotesize
$\frac{Sp(l+m)}{Sp(l)\times Sp(m)}\times \Z$
}
& 
 $Sp(n)$ & 
 $\frac{Sp(n)}{U(n)}$ & 
 $\frac{U(n)}{O(n)}$  \\ 
 \hline
 &  $p=0$ & $p=1$ & $p=2$ & $p=3$ & $p=4$ & $p=5$ & $p=6$ & $p=7$   \\ 
\hline
$d=0$ & $\Z$ & $\Z_2$ & $\Z_2$ & $0$ & $\Z$ & $0$ & $0$ & $0$ \\ 
$d=1$ & $0$ & $\Z$ & $\Z_2$ & $\Z_2$ & $0$ & $\Z$ & $0$ & $0$ \\
$d=2$ & $0$ & $0$ & $\Z$ & $\Z_2$ & $\Z_2$ & $0$ & $\Z$ & $0$ \\
$d=3$ & $0$ & $0$ & $0$ & $\Z$ & $\Z_2$ & $\Z_2$ & $0$ & $\Z$ \\
\hline
$d=4$ & $\Z$ & $0$ & $0$ & $0$ & $\Z$ & $\Z_2$ & $\Z_2$ & $0$ \\
$d=5$ & $0$ & $\Z$ & $0$ & $0$ & $0$ & $\Z$ & $\Z_2$ & $\Z_2$ \\
$d=6$ & $\Z_2$ & $0$ & $\Z$ & $0$ & $0$ & $0$ & $\Z$ & $\Z_2$ \\
$d=7$ & $\Z_2$ & $\Z_2$ & $0$ & $\Z$ & $0$ & $0$ & $0$ & $\Z$ \\
\hline
{\footnotesize
Example 
}
&
{\footnotesize
$\bmm
\text{insulator}\\
\text{w/ coplanar}\\
\text{spin order}
\emm$
}
&
{\footnotesize
$\bmm
\text{supercond.}\\
\text{w/ coplanar}\\
\text{spin order}
\emm$
}
&
{\footnotesize
$\bmm
\text{supercond.}
\emm$
}
&
{\footnotesize
$\bmm
\text{supercond.}\\
\text{w/ time}\\
\text{reversal}\\
\emm$
}
&
{\footnotesize
$\bmm
\text{insulator}\\
\text{w/ time}\\
\text{reversal}\\
\emm$
}
&
{\footnotesize
$\bmm
\text{insulator}\\[-1mm]
\text{w/ time}\\[-1mm]
\text{reversal and}\\[-1mm]
\text{inter-sublattice}\\[-1mm]
\text{hopping}
\emm$
}
&
{\footnotesize
$\bmm
\text{spin}\\
\text{singlet}\\
\text{supercond.}\\
\emm$
}
&
{\footnotesize
$\bmm
\text{spin}\\[-1mm]
\text{singlet}\\[-1mm]
\text{supercond.}\\[-1mm]
\text{w/ time}\\[-1mm]
\text{reversal}
\emm$
}
\\
\hline
 \end{tabular}
}
 \caption{
(Color online) Classification of gapped phases of non-interacting fermions in
$d$ spatial dimensions, for some symmetries.  The space of the gapped states is
given by $R_{p-d \text{ mod }8}$, where $p$ depends on the symmetry group.  The
phases are classified by $\pi_0(R_{p-d\text{ mod }8})$.  Here $\Z_2$ means that
there is one non-trivial phase and one trivial phase labeled by 1 and 0.  The
groups in red can be realized by electron systems (see Table \ref{elegrp}).
}
 \label{GFFsym}
\end{table*}

Since its introduction, we have been trying to obtain a systematic
understanding of topological orders. Some progresses have been made for certain
simple cases.  We found that all 2D Abelian topological orders can be
classified by integer $K$-matrices.\cite{BW9045,R9002,WZ9290}  The 2D
non-chiral topological orders (which can be smoothly connected to time reversal
and parity symmetric states) are classified by spherical fusion
category.\cite{FNS0428,LW0510,CGW1038,GWW1017} The recent realization of the
relation between topological order and long range entanglement\cite{CGW1038}
(defined through local unitary transformations\cite{VCL0501,V0705}) allows
us to separate another simple class of gapped quantum phases -- symmetry
protected topological (SPT) phases. SPT phases are gapped quantum phases with a
certain symmetry, which can all be smoothly connected to the same trivial
product state if we break the symmetry.  A generic construction of bosonic SPT
phases in any dimension using the group cohomology of the symmetry group was
obtained in \Ref{CLW1152,CGL1172}.  The constructed SPT phases include
interacting bosonic topological insulators and topological superconductors (and
much more).

Another type of simple systems are free fermion systems, for which a
classification of gapped quantum phases can be obtained through
K-theory\cite{K0886,SCR1101,AK1154} or non-linear $\si$-model of disordered
fermions.\cite{SRF0825}  They include the non-interacting topological
insulators\cite{KM0501,BZ0602,KM0502,MB0706,R0922,FKM0703,QHZ0824} and the
non-interacting topological
superconductors.\cite{SMF9945,RG0067,R0664,QHR0901,SF0904} Most gapped quantum
phases of free fermion systems are SPT phases protected by some symmetries,
such as topological insulators protected by the time reversal symmetry.  While
some others have intrinsic topological orders (\ie stable even without any
symmetry), such as topological superconductors with no symmetry.  Just like the
interacting topological ordered phases, the topological phases for free
fermions are also characterized by their gapless boundary excitations.  The
boundary excitations play a key role in the theory and experiments of free
fermion topological phases.

For non-interacting fermion systems with time reversal (generated by $\hat T$),
charge conjugation (generated by $\hat C$), and/or $U(1)$ (generated by $\hat
N$) symmetries, the total symmetry group may not simply be $Z_2^T\times
Z_2^C\times U(1)$. The group can take different forms, depending on the
different relations between those symmetry operations, such as $\hat T \hat N
\hat T^{-1}= \hat N$ or $\hat T \hat N \hat T^{-1}=- \hat N$.  As a result, the
gapped phases of those fermion systems with different symmetry groups have
different classifications.  In this paper, we use Kitaev's K-theory approach to
classify the gapped free fermion phases for those different symmetry groups.
In Table \ref{elegrp}, we list some electron systems and their full symmetry
group $G_f$.  In Table \ref{GFF} and  Table \ref{GFFsym} the 10
classes\cite{K0886,SRF0825} of gapped free fermion phases protected by those
many-body symmetry groups (and many other symmetry groups) are listed.  Here we
have assumed that the fermions form one irreducible representation of the full
symmetry group.  The result will differ if the fermions contain several
distinct irreducible representations of the full symmetry group (see section
\ref{gd}).  In \Ref{K0886,SRF0825}, the 10 classes of gapped free fermion
phases are already associated with many different many-body symmetries of
electron systems. In this paper, we generalize the results in
\Ref{K0886,SRF0825} to more symmetry groups.

We note that electron systems, with $\hat T \hat N \hat T^{-1}= \hat N$,
only realize a subset of the possible symmetry groups.  The emergent fermion
(such as the spinon in spin liquid) may realize other possible symmetry groups,
since their symmetries are described by projective symmetry groups (PSG) which
can be different for different topologically ordered states.\cite{Wqoslpub}

The $p=0$ line in Table \ref{GFF} classifies two types of electron systems:
(1) Insulators with only fermion number conservation (which includes integer
quantum Hall states). 
(2) Superconductors with only $S_z$ spin rotation symmetry which can be realized by superconductors with collinear spin order.  
The $p=1$ line in Table \ref{GFF} classifies 
superconductors with only time reversal and $S_z$
spin rotation symmetry (full symmetry group $G_-^+(U,T)$),
which can be realized by superconductors with real pairing
and $S_z$ conserving spin-orbital coupling.  

In Table \ref{GFFsym}, 
the $p=0$ column classifies electronic insulators with coplanar spin order
(full symmetry group $G_+^-(U,T)$ which contains the charge conservation and
a time reversal symmetry).
The $p=1$ column classifies electronic superconductors with coplanar spin order
and real pairing (full symmetry group $G_+(T)$ which contains a time reversal symmetry).
The $p=2$ column classifies electronic superconductors with non-coplanar spin
order (full symmetry group ``none'').
The $p=3$ column classifies electronic superconductors with spin-orbital
coupling and real pairing (full symmetry group $G_-(T)$ which contains the time reversal symmetry).
The $p=4$ column classifies electronic insulators with spin-orbital
coupling (full symmetry group $G_-^-(U,T)$ which contains the charge conservation and the time reversal symmetry).
The $p=5$ column classifies electronic insulators on bipartite lattices with
spin-orbital coupling and only inter-sublattice hopping (full symmetry group
$G_{--}^{-+}(U,T,C)$ which contains the charge conservation, the time reversal symmetry, and a charge conjugation symmetry).
The $p=6$ column classifies electronic spin singlet superconductors with
complex pairing (full symmetry group $SU(2)$).
The $p=7$ column classifies electronic spin-singlet superconductors with
real pairing (full symmetry group $G[SU(2),T]$ which contains
the $SU(2)$ spin rotation and the time reversal symmetry).


In this paper, we will first discuss a simpler case where fermion systems have
only the $U(1)$ symmetry.  Then we will discuss a more complicated case where
fermion systems can have time reversal, charge conjugation, and/or $U(1)$
symmetries. The classification of the gapped phases with translation symmetry
and the  classification of non-trivial defects with protected gapless
excitations will also be studied.

\section{Gapped free fermion phases -- the complex classes}

\subsection{The $d=0$ case}

Let us first consider a $0$-dimension free fermion system with $1$ orbital.
How many different gapped phases do we have for such a system? The answer is
$2$. The $2$ different gapped phases are labeled by $m=0,1$: the $m=0$ gapped
phase correspond to the empty orbital, while the $m=1$ gapped phase correspond
to the occupied orbital.  But $2$ is not the complete answer.  We can alway add
occupied and empty orbitals to the system and still regard the extended system
as in the same gapped phase.  So we should consider a system with $n$ orbitals
in $n\to \infty$ limit.  In this case, the $0$-dimension gapped phases are
labeled by an integer $m$ in $\Z$, where $m$ (with a possible constant shift) still corresponds to the number of
occupied orbitals.

Now let us obtain the above result using a fancier mathematical set up.  The
single-body Hamiltonian of the  $n$-orbital system is given by $n\times n$
hermitian matrix $H$.  If the orbitals below a certain energy are filled, we
can deform the energies of those orbitals to $-1$ and deform the energies of
other orbitals to $+1$ without closing the energy gap.  So, without losing the
generality, we can assume the $H$ to satisfy 
\begin{align}
 H^2=1
\end{align}
Such a hermitian matrix has a form
\begin{align}
 H=
U_{n\times n} 
\bpm 
I_{l\times l} & 0\\
0 &  -I_{m\times m}\\
\epm
U^\dag_{n\times n} 
\end{align}
where $n=l+m$ and $U_{n\times n} in U(n)$
is an $n\times n$ unitary matrix.
But $U_{n\times n}$ is not an one-to-one labeling of the
hermitian matrix satisfying $H^2=1$.
To obtain an one-to-one labeling, we note that
$\bpm 
I_{l\times l} & 0\\
0 &  -I_{m\times m}\\
\epm
$ is invariant under the unitary transformation
$\bpm 
V_{l\times l} & 0\\
0 &  W_{m\times m}\\
\epm
$
with $V_{l\times l} \in U(l)$ and
$W_{m\times m} \in U(m)$.
Thus the space $C_0$ of the hermitian matrix satisfying $H^2=1$
is given by
$\bigcup_m U(l+m)/U(l)\times U(m)$,
which, in $n\to \infty$ limit,
has a form
\begin{align}
 C_0\equiv \frac{U(l+m)}{U(l)\times U(m)} \times \Z
\end{align} 
Clearly $\pi_0(C_0)=\Z$, which recovers the result obtained above using a
simple argument: the $0$-dimension gapped phases of free conserved fermions are
labeled by integers $\Z$.  

\subsection{The properties of classifying spaces}

The space $C_0$ is the complex Grassmannian -- the space formed by the
subspaces of (infinity dimensional) complex vector space.  It is also the space
of the hermitian matrix satisfying $H^2=1$.  Actually, $C_0$ is a part of a
sequence.  
More generally, a space $C_p$ can be defined by first
picking $p$ fixed hermitian matrices $\ga_i$, $i=1,2,...,p$, satisfying
\begin{align}
\label{gaigaj}
\ga_i\ga_j+\ga_j\ga_i=2\del_{ij} .
\end{align}
Then $C_p$ is the space of the hermitian matrix satisfying 
\begin{align}
H^2=1,\ \ \ \ 
\ga_iH=-H\ga_i,\ \  i=1,...,p.  
\end{align}

To see what is the $C_1$ space,  let us choose $\ga_1=I_{n \times n}$ that
satisfy $\ga_1^2=1$.  But for such a choice, we cannot find any $H$ that
satisfy $\ga_1H=-H\ga_1$.  Actually, we have a condition on the choice of
$\ga_1$.  We must choose a $\ga_1$ such that $\ga_1H=-H\ga_1$ and $H^2=1$ has a
solution.  So we should choose  $\ga_1=\si^x\otimes I_{n \times n}$.  We note
that $\ga_1$ is invariant under the following unitary transformations $
\e^{\imth \si^x\otimes A_{n\times n}} \e^{\imth \si^0\otimes B_{n\times n}}\in
U(n)\times U(n) $ (where $A_{n\times n}$ and $B_{n\times n}$ are hermitian
matrices).  Then $H$ satisfying $H^2=1$ and $\ga_1H+H\ga_1=0$ has a form
\begin{align}
& H= 
\\
& \e^{\imth \si^x\otimes A_{n\times n}} \e^{\imth \si^0\otimes B_{n\times n}}
(\si^z\otimes I_{n\times n})
\e^{-\imth \si^0\otimes B_{n\times n}}
 \e^{-\imth \si^x\otimes A_{n\times n}},
\nonumber 
\end{align}
whose positive and negative eigenvalues are paired.
We see that the space $C_1$ is $U(n)\times U(n)/U(n)=U(n)$.

To construct the $C_2$ space, we can choose
$\ga_1=\si^x\otimes I_{n \times n}$ and
$\ga_2=\si^y\otimes I_{n \times n}$.
Then $H$ satisfying $H^2=1$ and $\ga_iH+H\ga_i=0$, $i=1,2$,  has a form
\begin{align}
 H= 
 \si^0\otimes U_{n\times n}
\left [\si^z\otimes \bpm
I_{l\times l} & 0\\
0 & -I_{m\times m}\\
\epm
\right]
\si^0\otimes U_{n\times n}^\dag
\end{align}
where $n=l+m$ and $ U_{n\times n} \in U(n)$.
We see that the space $C_2=C_0$.

To construct the $C_3$ space, we can choose
$\ga_1=\si^x\otimes I_{n \times n}$,
$\ga_2=\si^y\otimes I_{n \times n}$, and
$\ga_3=\si^z\otimes I_{n \times n}$.
But for such a choice, the equations
$\ga_iH+H\ga_i=0$, $i=1,2,3$, and $H^2=1$ has no solution for $H$.
So we need to 
impose the following condition on $\ga_i$'s:
\begin{align}
\label{condgai}
&\text{
The equations $\ga_iH+H\ga_i=0$, $H^2=1$
}
\nonumber \\
&\text{
have a solution for $H$.
}
\end{align}
(Later, we will see that such an condition has an amazing geometric
origin.) Let us choose
$\ga_1=\si^x\otimes \si^x \otimes I_{n \times n}$,
$\ga_2=\si^y\otimes \si^x \otimes I_{n \times n}$, and
$\ga_3=\si^z\otimes \si^x \otimes I_{n \times n}$ instead.
Then $H$ satisfying $H^2=1$ and $\ga_iH+H\ga_i=0$, $i=1,2,3$, has a form
\begin{align}
 H & = 
 \e^{\imth \si^0\otimes \si^x\otimes A_{n\times n}} \e^{\imth \si^0\otimes \si^0\otimes B_{n\times n}}
(\si^0\otimes \si^z\otimes I_{n\times n})\times
\nonumber\\
&\ \ \ \
\e^{-\imth \si^0\otimes \si^0\otimes B_{n\times n}}
 \e^{-\imth \si^0\otimes \si^x\otimes A_{n\times n}},
\end{align}
We find that $C_3=C_1$.

Now, it is not hard to see that $C_p=C_{p+2}$
which leads to $\pi_d(C_p)=\pi_{d}(C_{p+2})$.  
Thus
\begin{align}
 \pi_0(C_p)=
\begin{cases}
\Z, &  p=0 \text{ mod } 2, \\
\{ 0\}, &  p=1 \text{ mod } 2. \\
\end{cases}
\end{align}

\subsection{The $d\neq 0$ cases}

Next we consider a $d$-dimension free conserved fermion systems and their
gapped ground states. Note that the only symmetry that we have is the $U(1)$
symmetry associated with the fermion number conservation.  We do not have
translation symmetry and other symmetries.

To be more precise, our $d$-dimension space is a ball with no non-trivial
topology. Since the  systems have a boundary, here we can only require that the
``bulk'' gap of the fermion systems are non-zero. The free fermion system may
have protected gapless excited at the boundary.  (Requiring the fermion systems
to be even gapped at boundary will only give us trivial gapped phases.) We will
call the free  fermion systems that are gapped only inside of the $d$
dimensional ball as ``bulk'' gapped fermion systems.  A ``bulk'' gapped fermion
system may or may not be gapped at the boundary.

Kitaev has shown that the space $C^H_d$ of such ``bulk'' gapped free fermion
systems is homotopically equivalent of the space $C^M_d$ of mass matrices of a
$d$-dimensional Dirac equation: $\pi_n(C^H_d)=\pi_n(C^M_d)$.\cite{K0886}  In
the following, we will give a hand-waving explanation of the result.

To start, let us first assume that the fermion system has the translation
symmetry and the charge conjugation symmetry. We also assume that its energy
bands have some Dirac points at zero energy and there are no other zero energy
states in the Brillouin zone.  So if we fill the negative energy bands, the
single-body gapless excitations in the system are described by the hermitian
matrix $H$, whose continuous limit has a form:
\begin{align}
 H = \sum_{i=1}^d \ga_i \imth\prt_i ,
\end{align}
where we have folded all the Dirac points to the $\v k=0$ point.  Without
losing generality, we have also assumed that all the Dirac points have the same
velocity.  Since $\imth\prt_i$ is hermitian, thus $\ga_i$, $i=1,...,d$ are the
hermitian $\ga$-matrices (of infinity dimension) that satisfy \eqn{gaigaj}.

When $d=1$, do we have a system that has $\ga_1=I_{n\times n}$?
The answer is no. Such a system will have $n$ right-moving chiral modes
that cannot be realized by any pure $1$-dimensional systems
with short-ranged hopping.
In fact $\ga_1$ must have a form $\ga_1=
\bpm
I_{l\times l} & 0\\
0 & -I_{m\times m}\\
\epm
$ with $l=m$ (the same number of right-moving and left-moving modes).  So the
allowed $\ga_1$ always satisfy the condition that $H^2=1$ and $\ga_1H+H\ga_1=0$
has a solution for $H$.  We see that the extra condition \eqn{condgai} on
$\ga_i$ has a very physical meaning.

Now we add perturbations that may break the translation 
symmetry. We like to know
how many different ways are there to gap the Dirac point.
The Dirac points can be fully gapped by the ``mass'' matrix $M$ that
satisfies
\begin{align}
 \ga_i M + M\ga_i=0,\ \ \ \ M^\dag=M.
\end{align}
To fully gap the Dirac points, $M$ must have no zero eigenvalues.
Without losing generality, we may also assume that
$ M^2=1$.
(Since the Dirac points may have different crystal momenta before the folding
to $\v k=0$, we need perturbations that break the translation symmetry to
generate a generic mass matrix that may mix Dirac points at different crystal
momenta.) The space $C_d^M$ of such mass matrices is nothing but $C_d$
introduced before: $  C_d^M=C_d$.

So the different ways to gap the Dirac point form a space $C_{d}$.  The
different disconnected components of $C_{d}$ represent different gapped phases
of the free fermions.  Thus, the gapped phases of the conserved free fermions
in $d$-dimensions are classified by $\pi_0(C_{d})$, which is $\Z$ for even $d$
and $0$ for odd $d$.  The non-trivial phases at $d=2$ are labeled by $\Z$,
which are the integer quantum Hall states.  The results is summarized in Table
\ref{GFF}.

\section{Gapped free fermion phases 
-- the real classes}

When fermion number is not conserved and/or when there is a time reversal
symmetry, the gapped phases of non-interacting fermions are classified
differently. However, using the idea and the approaches similar to the above
discussion, we can also obtain a classification.  Instead of considering
hermitian matrices that satisfy certain conditions, we just need to consider
real antisymmetric matrices that satisfy certain conditions.  

\subsection{The $d=0$ case -- the symmetry groups}

Again, we will start with $d=0$ dimension.  
In this case, a free fermion system with $n$ orbitals is described by the
following quadratic Hamiltonian
\begin{align}
 \hat H= 
\sum_{ij} H_{ij} \hat c_i^\dag \hat c_j
+\sum_{ij} [G_{ij} \hat c_i \hat c_j+ h.c.], \ \ \ 
\ i,j=1,...,n
\end{align}
Introducing
Majorana fermion operator $\hat \eta_I$, $I=1,...,2n$,
\begin{align}
 \{\hat\eta_I,\hat\eta_J\}=2\del_{IJ},\ \ \ \ \ \hat\eta_I^\dag=\hat\eta_I,
\end{align}
to express the complex fermion operator $\hat c_i$:
\begin{align}
 \hat c_i=\frac12( \hat \eta_{2i}+\imth \eta_{2i+1} ),
\end{align}
we can rewrite $\hat H$ as
\begin{align}
\hat  H=\frac{\imth}{4}\sum_{IJ} A_{IJ}\hat\eta_I\hat\eta_J.
\end{align}
where $A$ is a real antisymmetric matrix.
For example, for a 1-orbital Hamiltonian
$\hat H=\eps (\hat c^\dag \hat c-\frac12)$, we get
$A=\bpm
0& -\eps\\
\eps& 0\\
\epm$.

If the fermion number is conserved, $\hat H$ commutes with the fermion number
operator 
\begin{align}
\hat N\equiv \sum_i(\hat c_i^\dag \hat c_i-\frac12)
=\frac{\imth}{4}\sum_{IJ} Q_{IJ}\hat\eta_I\hat\eta_J
\end{align}
where
\begin{align}
 Q=\veps \otimes I,\ \ \ \ Q^2=-1, \ \ \ \ \veps\equiv -\imth\si^y.
\end{align}
$[\hat H,\hat N]=0$ requires that
\begin{align}
 [A,Q]=0.
\end{align}
Such matrix $A$ has a form $A=\si^0\otimes H_a + \veps \otimes H_s$, where
$H_s$ is symmetric and  $H_a$ antisymmetric.  We can convert such an
antisymmetric matrix $A$ into a hermitian matrix $H=H_s+\imth H_a$, and reduce
the problem to the one discussed before.

\begin{table*}[tb]
 \centering
 \begin{tabular}{ |c|c| }
 \hline
Symmetry groups  & Relations \\
\hline
\hline
$U(1),\ SU(2)$ & 
\\
\hline
$G_{s_C}(C)$ & 
$\hat C^2 = s_C^{\hat N}$,~~ $s_C=\pm $.
\\
\hline
$G_{s_C}(U,C)$ & 
$\hat C^2 = s_C^{\hat N}$,~ 
$\hat C \e^{\imth \th \hat N} \hat C^{-1} = \e^{-\imth \th \hat N}$,~
$s_C=\pm$.
\\
\hline
$G_{s_T}(T)$ & 
$\hat T^2 = s_T^{\hat N}$,~~ $s_T=\pm$.
\\
\hline
$G_{s_T}^{s_{UT}}(U,T)$ & 
$\hat T \e^{\imth \th \hat N} \hat T^{-1} = \e^{s_{UT}\imth \th \hat N}$,~
$\hat T^2 = s_T^{\hat N}$,~~ $s_{UT},s_T=\pm$.
\\
\hline
$G_{s_Ts_C}^{s_{TC}}(T,C)$ & 
$\hat T^2 = s_T^{\hat N}$,~ 
$\hat C^2 = s_C^{\hat N}$,~ 
$\hat C\hat T = (s_{TC}^{\hat N}) \hat T\hat C$,~~ 
$s_{TC},s_T,s_C=\pm$.
\\
\hline
$G_{s_Ts_C}^{s_{UT}s_{TC}}(U,T,C)$ & 
$\hat C \e^{\imth \th \hat N} \hat C^{-1} = \e^{-\imth \th \hat N}$,~
$\hat T \e^{\imth \th \hat N} \hat T^{-1} = \e^{s_{UT}\imth \th \hat N}$,~
\\
&
$\hat T^2 = s_T^{\hat N}$,~ 
$\hat C^2 = s_C^{\hat N}$,~ 
$\hat C\hat T = (s_{TC}^{\hat N}) \hat T\hat C$,~~ 
$s_T,s_C,s_{UT},s_{TC}=\pm$.
\\
\hline
 \end{tabular}
\caption{
Different relations between symmetry transformations gives rise to 36 different
groups that contain $U(1)$ (represented by $U$), time reversal $T$, and/or
charge conjugation $C$ symmetries.
}
 \label{groups}
\end{table*}

The time reversal transformation $\hat T$ is antiunitary: $\hat T \imth \hat
T^{-1}=-\imth$.  Since $\hat T$ does not change the fermion numbers, thus $\hat
T \hat c_i \hat T^{-1} = U_{ij}\hat c_j$, where $U$ is an unitary matrix.  
In terms of the Majorana fermions, we have
\begin{align}
 \hat T \hat \eta_{2i} \hat T^{-1} &= 
\Re{U_{ij}}\hat \eta_{2j} -\Im{U_{ij}}\hat \eta_{2j+1},
\nonumber\\
 \hat T \hat \eta_{2i+1} \hat T^{-1} &= 
-\Re{U_{ij}}\hat \eta_{2j+1} -\Im{U_{ij}}\hat \eta_{2j}.
\end{align}
Therefore, we have
\begin{align}
 \hat T \hat \eta_{i} \hat T^{-1} &= 
T_{ij}  \hat \eta_{j},\ \ \
T=\si^3\otimes \Re U - \si^1\otimes \Im U .
\end{align}
We see that in the
Majorana fermion basis, $U\to \si^3\otimes \Re U - \si^1\otimes \Im U=T$
and $\imth \to \eps\otimes I$.
We indeed have $T (\eps\otimes I) =-T (\eps\otimes I)$.

For
fermion systems, we may have $\hat T^2=s_T^{\hat N}$ $s_T=\pm$. In fact
$s_T=-$ for electron systems.  This implies that $\hat T^2 \hat c_i \hat T^{-2} =s_T
\hat c_i$ and $T^2=s_T$.  
The time reversal invariance $\hat T\hat H\hat T^{-1}=\hat H$ implies that
$ T^\top A T = -A$, where $T^\top$ is the transpose of $T$.  We can show that
\begin{align}
 T^\top T &=
(\si^3\otimes \Re U^\dag - \si^1\otimes \Im U^\dag)
(\si^3\otimes \Re U + \si^1\otimes \Im U)
\nonumber\\
&=\ \si^0 \otimes (\Re U^\dag \Re U- \Im U^\dag \Im U)
\nonumber\\
&\ \ \ \, - \eps \otimes (\Re U^\dag \Im U + \Im U^\dag \Re U)
\nonumber\\
&=\si^0 \otimes I,
\end{align}
where we have used
$\Re U^\dag \Re U- \Im U^\dag \Im U=I$ and
$\Re U^\dag \Im U + \Im U^\dag \Re U=0$ 
for unitary matrix $U$.
Therefore $T^\top =T^{-1}$ and
\begin{align}
\label{Tinv}
 A T = - TA,\ \ \ \  T^2=s_T .
\end{align}
Also, for fermion systems, the time reversal transformation $\hat T$ and the
$U(1)$ transformation $\hat N$ may have a non-trivial relation: $\hat T
\e^{\imth \th \hat N}\hat T^{-1}= \e^{s_{UT}\imth \th \hat N}, \ s_{UT}=\pm $,
or $ \hat T \hat N\hat T^{-1} = -s_{UT} \hat N $.  This gives us
\begin{align}
TQ= s_{UT}QT .
\end{align}

The charge conjugation transformation $\hat C$ is unitary.  Since $\hat C$
changes $\hat c_i$ to $\hat c_i^\dag$, thus $\hat C \hat c_i \hat C^{-1} =
U_{ij}\hat c_j^\dag$, where $U$ is an unitary matrix.  
In terms of the Majorana fermions, we have
\begin{align}
 \hat C \hat \eta_{2i} \hat C^{-1} &= 
\Re{U_{ij}}\hat \eta_{2j} +\Im{U_{ij}}\hat \eta_{2j+1},
\nonumber\\
 \hat C \hat \eta_{2i+1} \hat C^{-1} &= 
-\Re{U_{ij}}\hat \eta_{2j+1} +\Im{U_{ij}}\hat \eta_{2j}.
\end{align}
Therefore, we have
\begin{align}
 \hat C \hat \eta_{i} \hat C^{-1} &= 
C_{ij}  \hat \eta_{j},\ \ \
C=\si^3\otimes \Re U + \si^1\otimes \Im U .
\end{align}
Again, we can show that $C^\top=C^{-1}$.


For fermion systems, we may have $\hat C^2=s_C^{\hat N}, \
s_C=\pm$, which implies that $\hat C^2 \hat c_i \hat C^{-2} =s_C\hat c_i$ and
$C^2=s_C$.  
%
The
charge conjugation invariance $\hat C\hat H\hat C^{-1}=\hat H$ implies that
$A$ satisfies
\begin{align}
\label{Cinv}
 C A  = CA,\ \ \ \  C^2=s_C .
\end{align}
Since $\hat C \hat N\hat C^{-1}=-\hat N$, we have
\begin{align}
 CQ=-QC .
\end{align}
However, the commutation relation
between $\hat T$ and $\hat C$ has two choices:  
$\hat T\hat C=s_{TC}^{\hat N}\hat C\hat T, \ s_{TC}=\pm$, we have
\begin{align}
& CT=s_{TC}TC.
\end{align}

We see that when we say a system has $U(1)$, time reversal, and/or charge
conjugation symmetries, we still do not know what is the actual symmetry group
of the system, since those symmetry operations may have different relations as
described by the signs $s_T,s_C,s_{UT},s_{TC}$, which lead to different full
symmetry groups.  Because symmetry plays a key role in our classification, we
cannot obtain a classification without specifying the symmetry groups.  We have
discussed the possible relations among various symmetry operations. In the
Table \ref{groups}, we list the corresponding symmetry groups.

We like to point that that some times, when we describe the symmetry of a
fermion system, we do not include the fermion number parity transformation
$(-)^{\hat N}$ in the symmetry group $G$.  However, in this paper, we will use
the full symmetry group $G_f$ to describe the symmetry of a fermion system. The
full symmetry group $G_f$ does include the fermion number parity transformation
$(-)^{\hat N}$.  So the full symmetry group of a fermion system with no
symmetry is $G_f=Z_2^f$ generated by the  fermion number parity transformation.
$G_f$ is actually a $Z_2^f$ extension of $G$: $G=G_f/Z_2^f$.  It is a
projective symmetry group discussed in \Ref{Wqoslpub}.

In the following, we will study the symmetries of various electron systems, 
to see which symmetry groups listed in Table \ref{groups} can be realized by
electron systems.  

For insulators with non-coplanar spin order $\del H= \hat
c_i^\dag \v n_1 \cdot \v \si \hat c_i + \hat c_j^\dag \v n_2 \cdot \v \si \hat
c_j + \hat c_k^\dag \v n_3 \cdot \v \si \hat c_k $, the full symmetry group is
$G_f=U(1)$ generated by the total charge $\hat N_C$.

For superconductors  with non-coplanar spin order $ \del H= \hat c_i^\dag \v
n_1 \cdot \v \si \hat c_i + \hat c_j^\dag \v n_2 \cdot \v \si \hat c_j + \hat
c_k^\dag \v n_3 \cdot \v \si \hat c_k +(\hat c_{\up i}\hat c_{\down j}- \hat
c_{\down i}\hat c_{\up j}) $, the full symmetry group is reduced to $G_f=Z_2^f$
generated by the fermion number parity operator $P_f=(-)^{\hat N_C}$.  We note
that the full symmetry group of any fermion system contain $Z_2^f$ as a
subgroup. So we usually, use the group $G_f/Z_2^f$ to describe the symmetry of
fermion system, and we say there is no symmetry for superconductors  with
non-coplanar spin order.  But in this paper, we will use  the full symmetry
group $G_f$ to describe the symmetry of fermion systems.

For insulators with spin-orbital coupling $\del H= \imth \hat c_i^\dag \v n_1
\cdot \v \si \hat c_{i'} +\imth \hat c_j^\dag \v n_2 \cdot \v \si \hat c_{j'}
+\imth \hat c_k^\dag \v n_3 \cdot \v \si \hat c_{k'} $, they have the charge
conservation $\hat N_C$ and the time reversal $\hat T_\text{phy}$ symmetries.
The time reversal symmetry is defined by
\begin{align}
\hat T_\text{phy} \hat c_{\al,i} \hat T_\text{phy}^{-1}= \eps_{\al\bt} \hat c_{\bt,i},\ \ \
\hat T_\text{phy} \hat c^\dag_{\al,i} \hat T_\text{phy}^{-1}= \eps_{\al\bt} \hat c^\dag_{\bt,i}.
\end{align}
We can show that
\begin{align}
& \hat T_\text{phy} \hat c_i^\dag \v \si \hat c_j \hat T_\text{phy}^{-1}
=- \hat c_i^\dag \v \si \hat c_j,
\nonumber\\
&
 \hat T_\text{phy} \hat N_C \hat T_\text{phy}^{-1}
= \hat N_C,\ \ \
\hat T^2_\text{phy} = (-)^{\hat N_C}
\end{align}
Thus $\del H$ is invariant under
$\hat T_\text{phy} $ and $\e^{\imth \th \hat N_C}$.
Let 
$\hat T=\hat T_\text{phy} $ and $\hat N= \hat N_C$,
we find that
\begin{align}
\hat T \e^{\imth \th \hat N} \hat T^{-1} = \e^{-\imth \th \hat N},
\ \ \ 
\hat T^2 = (-)^{\hat N}
,
\end{align}
which define the full symmetry group $G_-^-(U,T)$ of an electron insulator
with spin-orbital coupling.

For superconductors with spin-orbital coupling and real pairing
$\del H=\imth \hat c_i^\dag \v n_1 \cdot \v \si \hat c_{i'}
+\imth \hat c_j^\dag \v n_2 \cdot \v \si \hat c_{j'}
+\imth \hat c_k^\dag \v n_3 \cdot \v \si \hat c_{k'}
+(\hat c_{\up i}\hat c_{\down j}- \hat c_{\down i}\hat c_{\up j})$,
they have the time reversal symmetry $\hat T_\text{phy}$.
Setting
$\hat T=\hat T_\text{phy} $ and $\hat N= \hat N_C$,
we find
\begin{align}
\hat T^2 = (-)^{\hat N}
,
\end{align}
which defines the full symmetry group $G_-^+(U,T)=Z_4$
of superconductors with spin-orbital coupling and real pairing.
 
For superconductors with $S_z$ conserving spin-orbital coupling and real pairing
$\del H=\imth \hat c_i^\dag \si^z \hat c_{j}
+(\hat c_{\up i}\hat c_{\down j}- \hat c_{\down i}\hat c_{\up j})$,
they have the time reversal $\hat T_\text{phy}$ and
$S_z$ spin rotation symmetries.
Setting
$\hat T=\hat T_\text{phy} $ and $\hat N= 2\hat S_z$,
we find
\begin{align}
\hat T \e^{\imth \th \hat N} \hat T^{-1} = \e^{\imth \th \hat N},
\ \ \ 
\hat T^2 = (-)^{\hat N}
,
\end{align}
which defines the full symmetry group $G_-^+(U,T)=U(1)\times Z_2$
of superconductors with $S_z$ conserving spin-orbital coupling and real pairing.

For superconductors with real pairing and coplanar spin order $\del H= \hat
c_i^\dag \v n_1 \cdot \v \si \hat c_i +\hat c_j^\dag \v n_2 \cdot \v \si \hat
c_j +(\hat c_{\up i}\hat c_{\down j}- \hat c_{\down i}\hat c_{\up j})$, they
have a combined time reversal and $180^\circ$ spin rotation symmetry.  The spin
rotation is generated by $S_a =\sum_i \frac12 c_i^\dag  \si^a c_i$, $a=x,y,z$.
We have
\begin{align}
\hat T_\text{phy} \hat S_a \hat T_\text{phy}^{-1}= -  \hat S_a.
\end{align}
The Hamiltonian $\del H$ is invariant under $\hat T=  \e^{\imth \pi \hat
S^y}\hat T_\text{phy}$.  Since $\hat T^2=1$, the full symmetry group of
superconductors real pairing and coplanar spin order is $G_+(T)=Z_2\times
Z_2^f$.

For superconductors with real pairing and collinear spin order $\del H=\hat
c_i^\dag\si^z \hat c_j +(\hat c_{\up i}\hat c_{\down j}- \hat c_{\down i}\hat
c_{\up j})$, they have the $S_z$ spin rotation and a combined time reversal and
$180^\circ$ $S_y$ spin rotation symmetry.
The Hamiltonian $\del H$ is invariant under $\hat T=  \e^{\imth \pi \hat
S^y}\hat T_\text{phy}$ and $S_z$ spin rotation $\hat N=2\hat S_z$.
We find that
\begin{align}
\hat T \e^{\imth \th \hat N} \hat T^{-1} = \e^{-\imth \th \hat N},
\ \ \ 
\hat T^2 = (-)^{\hat N}
,
\end{align}
which define the full symmetry group $G_+^-(U,T)$ of 
superconductors with real pairing and collinear spin order.

For insulators with coplanar spin order $ \hat c_i^\dag \v n_1 \cdot \v \si
\hat c_i +\hat c_j^\dag \v n_2 \cdot \v \si \hat c_j $, they have the charge
conservation and a combined time reversal and $180^\circ$ spin rotation
symmetries.
The Hamiltonian $\del H$ is invariant under $\hat T=  \e^{\imth \pi \hat
S^y}\hat T_\text{phy}$ and the charge rotation $\hat N=\hat N_C$.
We find that
\begin{align}
\hat T \e^{\imth \th \hat N} \hat T^{-1} = \e^{-\imth \th \hat N},
\ \ \ 
\hat T^2 = (-)^{\hat N}
,
\end{align}
which define the full symmetry group $G_+^-(U,T)$ of 
insulators with coplanar spin order.

For superconductors with real triplet $S_z=0$ paring $\del H=\hat c_{\up i}\hat
c_{\down j}+ \hat c_{\down i}\hat c_{\up j}$, they have 
a combined time reversal and charge rotation symmetry, 
a combined $180^\circ$ $S_y$ spin rotation and charge rotation symmetry, 
and the $S_z$-spin rotation symmetry.
The Hamiltonian $\del H$ is invariant under 
$\hat T=  \e^{\imth \frac{\pi}{2} \hat N_C}\hat T_\text{phy}$, 
$\hat C=  \e^{\imth \frac{\pi}{2} \hat N_C} \e^{\imth \pi \hat S_y}$, 
and the $S_z$ spin rotation $\hat N=2\hat S_z$.
We find that
\begin{align}
&
\hat C \e^{\imth \th \hat N} \hat C^{-1} = \e^{-\imth \th \hat N},
\nonumber\\
&
\hat T \e^{\imth \th \hat N} \hat T^{-1} = \e^{\imth \th \hat N},
\nonumber\\
&
\hat T^2 = 1, \ \ \
\hat C^2 = 1,\ \ \
\hat C\hat T = (-)^{\hat N} \hat T\hat C 
\end{align}
which define the full symmetry group $G_{++}^{--}(U,T,C)$ of 
superconductors with real triplet $S_z=0$ paring.

For superconductors with real triplet $S_z=0$ paring 
and collinear spin order $\del H=\hat c_i \si^z c_i+(\hat c_{\up i}\hat
c_{\down j}+ \hat c_{\down i}\hat c_{\up j})$, they have 
a combined time reversal and $180^\circ$ $S_y$ spin rotation symmetry, 
and the $S_z$-spin rotation symmetry.
The Hamiltonian $\del H$ is invariant under 
$\hat T=  \e^{\imth \pi \hat S_y} \hat T_\text{phy}$, 
and the $S_z$ spin rotation $\hat N=2\hat S_z$.
We find that
\begin{align}
&
\hat T \e^{\imth \th \hat N} \hat T^{-1} = \e^{-\imth \th \hat N},
\nonumber\\
&
\hat T^2 = 1, \ \ \
\end{align}
which define the full symmetry group $G_{+}^{-}(U,T)$ of 
superconductors with real triplet $S_z=0$ paring 
and collinear spin order.


For superconductors with the time reversal, the $180^\circ$ $S_y$-spin rotation,
and the $S_z$-spin rotation symmetries, the Hamiltonian is invariant under
$\hat T=  \hat T_\text{phy}$, $\hat C=  \e^{\imth \pi \hat S_y}$, and $\hat
N=2\hat S_z$.
We find that
\begin{align}
& \hat C \e^{\imth \th \hat N} \hat C^{-1} = \e^{-\imth \th \hat N},\ \ 
\hat T \e^{\imth \th \hat N} \hat T^{-1} = \e^{\imth \th \hat N},
\nonumber\\
&
\hat T^2 = \hat C^2 = (-)^{\hat N}, \ \ 
\hat C\hat T = \hat T\hat C  .
\end{align}
which define the full symmetry group $G_{--}^{++}(U,T,C)$ of superconductors
with the time reversal, the $180^\circ$ $S_y$-spin rotation, and the $S_z$-spin
rotation symmetries.  For free electrons with the $180^\circ$ $S_y$-spin
rotation, and the $S_z$-spin rotation symmetries, they actually have the full
$SU(2)$ spin rotation symmetry.  So the above systems are also superconductors
with real pairing and the $SU(2)$ spin rotation symmetry.  Similarly, for
superconductors with complex pairing and the $SU(2)$ spin rotation symmetry,
the symmetry group is $SU(2)$, or $G_-(U,C)$.

For insulators with spin-orbital coupling and only inter-sublattice hopping $H=
\imth \hat c_{i_A}^\dag \v n_1 \cdot \v \si \hat c_{i_B} +\imth \hat
c_{j_A}^\dag \v n_2 \cdot \v \si \hat c_{j_B} +\imth \hat c_{k_A}^\dag \v n_3
\cdot \v \si \hat c_{k_B} $, they have the charge conservation, the time
reversal and a deformed charge conjugation symmetries.  The charge conjugation
transformation $\hat C_\text{phy}$ is defined as
\begin{align}
\hat C_\text{phy} \hat c_{\al,i} \hat C_\text{phy}^{-1}= \eps_{\al\bt} \hat c^\dag_{\bt,i},\ \ \
\hat C_\text{phy} \hat c^\dag_{\al,i} \hat C_\text{phy}^{-1}= \eps_{\al\bt} \hat c_{\bt,i}.
\end{align}
We find that
\begin{align}
& \hat C_\text{phy} \hat c_i^\dag \v \si \hat c_j \hat C_\text{phy}^{-1}
= \hat c_i^\dag \v \si \hat c_j, \ \ \
\hat C_\text{phy}
\hat T_\text{phy}
=
\hat T_\text{phy}
\hat C_\text{phy},
\nonumber\\
&
 \hat C_\text{phy} \hat N_C \hat C_\text{phy}^{-1}
= -\hat N_C,\ \ \
\hat C^2_\text{phy} = (-)^{\hat N_C}
\end{align}
The above Hamiltonian is invariant under $\hat T=\hat T_\text{phy}$, $\hat
N=\hat N_C$, and $\hat C= (-)^{\hat N_B} \hat C_\text{phy}$, where $\hat N_B$
is the number of electrons on the $B$-sublattice.
We find
\begin{align}
&
\hat C \e^{\imth \th \hat N} \hat C^{-1} = \e^{-\imth \th \hat N},
\nonumber\\
&
\hat T \e^{\imth \th \hat N} \hat T^{-1} = \e^{-\imth \th \hat N},
\nonumber\\
&
\hat T^2 = (-)^{\hat N}, \ \ \
\hat C^2 = (-)^{\hat N},\ \ \
\hat C\hat T = \hat T\hat C 
\end{align}
which define the full symmetry group $G_{--}^{-+}(U,T,C)$ of
insulators with spin-orbital coupling and only inter-sublattice hopping.
The above results for electron systems and their full symmetry groups $G_f$
are summarized in Table \ref{elegrp}.

\subsection{The $d=0$ case -- the classifying spaces}

The hermitian matrix $\imth A$ describes single-body excitations above the free
fermion ground state.  We note that the eigenvalues of $\imth A$ are $\pm
\eps_i$.  The positive eigenvalues $|\eps_i|$ correspond to the single-body
excitation energies above the many-body ground state.  The minimal $|\eps_i|$
represents the excitation energy gap of $\hat H$ above the ground state (the
ground state is the lowest energy state of  $\hat H$).  So, if we are
considering gapped systems, $|\eps_i|$ is alway non-zero. We can shift all
$\eps_i$ to $\pm 1$ without closing the gap and change the (matter) phase of
the state.  Thus, we can set $A^2=-1$.

In presence of symmetry, $A$ should also satisfy some other conditions.  The
space formed by all those $A$'s is called the classifying space.  Clearly, the
classifying space is determined by the full symmetry group $G_f$.  In this
section, we will calculate the classifying spaces for some simple groups.

If there is no symmetry, then the real antisymmetric matrix
$A$ satisfies
\begin{align}
 A^2=-1.
\end{align}
The space of those matrices is denoted as $R^0_0$,
which is the classifying space for trivial symmetry group.

If there is only the charge conjugation symmetry
(full symmetry group = $G_{s_C}(C)$), 
then the real antisymmetric matrix
$A$ satisfies
\begin{align}
 A^2=-1,\ \ \ \ AC=CA,\ \ \ \ C^2=s_C.
\end{align}
For $s_C=+$, since $C$ commute with $A$ and $C$ is symmetric, we can always
restrict ourselves in an eigenspace of $C$ and $C$ can be dropped.  Thus the
space of the matrices is $R^0_0$, the same as before.  For $s_C=-$, we can
assume $C=\veps\otimes I$. In this case, $A$ has a form $A=\si^0\otimes
H_a+\veps\otimes H_s$, where $H_s=H_s^T$ and $H_a=-H_a^T$.  Thus, we can
convert $A$ into a hermitian matrix $H=H_s+\imth H_a$, and the space of the
matrices is $C_0$.

If there are $U(1)$ and charge conjugation symmetries
(full symmetry group = $G_{s_C}(U,C)$), 
then the real antisymmetric matrix
$A$ satisfies
\begin{align}
& A^2=-1,\ \
AQ=QA,\ \
AC=CA,\ \
QC=-CQ,   
\nonumber\\
&
Q^2=-1, \ \
C^2=s_C.
\end{align}
For $s_C=+$, we can assume $C=\si^z\otimes I$ and $Q=\veps\otimes I$.  Since
$Q$ and $C$ commute with $A$, we find that $A$ must have a form $A=\si^0\otimes
\t A$, with $\t A^2=-1$.  Thus the space of the matrices $\t A$, and hence $A$,
is $R^0_0$.  

For $s_C=-$, we can assume $C=\veps\otimes \si^x\otimes I$ and $Q=\veps\otimes
\si^z\otimes I$.  We find that $A$ must have a form $A=\si^0\otimes\si^0
\otimes H_0 +\veps\otimes\si^0 \otimes H_1 +\si^z\otimes\veps \otimes H_2
+\si^x\otimes\veps \otimes H_3 $, where $H_0=-H_0^T$ and $H_i=H_i^T$,
$i=1,2,3$.  Now, we can view $\si^0\otimes\si^0$ as $1$, $\veps\otimes\si^0$ as
i, $\si^z\otimes\veps$ as j, and $\si^x\otimes\veps$ as k.  We find that i,j,k
satisfy the quaternion algebra.  Thus $A$ can be mapped into a quaternion
matrix $H=H_0 +\text{i}H_1 +\text{j}H_2 +\text{k}H_3$ satisfying $H^\dag=-H$
and $H^2=-1$.  The quaternion matrices that satisfy the above two conditions
has a form
\begin{align}
H=\e^{X_{n\times n}} \imth I_{n\times n} \e^{-X_{n\times n}} 
\end{align}
where $X^\dag_{n\times n}=-X_{n\times n}$ is a quaternion matrix.
$\e^{X_{n\times n}}$ form the group $Sp(n)$.  However, the transformations
$\e^{A_{n\times n}+\imth B_{n\times n}}$ keeps $\imth I_{n\times n}$ unchanged,
where $A_{n\times n}$ is a real antisymmetric matrix and $B_{n\times n}$ is a
real symmetric matrix.  $\e^{A_{n\times n}+\imth B_{n\times n}}$ form the group
$U(n)$.  Thus, the space of the quaternion matrices that satisfy the above two
conditions is given by $Sp(n)/U(n)$.  Such a space is the space $R_6$ which
will be introduced later.

When $s_C=-1$, we can view $Q$ as the generator of $\hat S_z$ spin rotation,
and $C$ as the generator of $\hat S_x$ spin rotation acting on spin-1/2
fermions.  In fact $\e^{\th_z Q}$ and $\e^{\th_x C}$ in this case generate the
full $SU(2)$ group.  So when $s_C=-$, the free spin-1/2 fermions with
$U(1)\rtimes Z_2^C$ symmetry actually have the full $SU(2)$ spin rotation
symmetry.  Therefore $G_{-}(U,C) \sim SU(2)$.

If there is only the time reversal symmetry (full symmetry group =
$G_{s_T}(T)$), then $A$ satisfies
\begin{align}
 A^2=-1, \ \ \ \ A\rho_1+\rho_1 A=0,\ \ \ \ \rho_1^2=s_T, \ \ \ \
\rho_1=T.
\end{align}
The space of those matrices is denoted as $R^1_0$ for $s_T=-1$ and $R^0_1$ for
$s_T=1$.

If there are time reversal and $U(1)$ symmetries
(full symmetry group = $G_{s_T}^{s_{UT}}(U,T)$), then for
$s_{UT}=-$, $A$ satisfies
\begin{align}
&
 A^2=-1, \ \ A\rho_i+\rho_i A=0,\ \ 
\rho_1^2=s_T, \ \ 
\rho_2^2=s_T, 
\nonumber\\
&
\rho_1=T, \ \ \rho_2=TQ.
\end{align}
The space of those matrices is denoted as $R^0_2$ for $s_T=+$ and $R^2_0$ for
$s_T=-$.  

For $s_{UT}=+$, $Q$ commute with both $A$ and $T$.  Since $Q^2=-1$, we can
treat $Q$ as the imaginary number $\imth$ and convert both $A$ and $T$ to
complex matrices.  To see this, let us choose a basis in which $Q$ has a form
$Q=\eps\otimes I$.  In this basis $A$ and $T$ become $A=\si^0\otimes A_2 + \eps
\otimes A_1$ and $T=\si^0\otimes T_1 + \eps \otimes T_2$, where $A_1$ is
symmetric and $A_2$ is antisymmetric.  
Let us introduce complex matrices $H=-A_1+\imth A_2$ and  
$\t T=T_1+\imth T_2$ for $s_T=+$ or
$\t T=-T_2+\imth T_1$ for $s_T=-$.
From $A^2=-1$, $T^2=s_T$, and $AT=-TA$, we find
\begin{align}
H^2=1,\ \ \ H\t T+\t T H=0,\ \ \ \ \t T^2=1.
\end{align}
Also $A^T=-A$ allows us to show $H^\dag =H$.  For a fixed $\t T$, the space
formed by $H$'s that satisfy the above conditions is $C_1$ introduced before.
This allows us to show that the space of the corresponding matrices $A$ is
$C_1$ for $s_T=\pm $, $s_{UT}=+$.

If there are time reversal and
charge conjugation symmetries
(full symmetry group = $G_{s_Ts_C}^{s_{TC}}(T,C)$), then for
$s_{TC}=-$, $A$ satisfies
\begin{align}
&
 A^2=-1, \ \  A\rho_i+\rho_i A=0,\ \  
\rho_1^2=s_T, \ \ 
\rho_2^2=-s_Ts_C, 
\nonumber\\
&
\rho_1=T, \ \ \rho_2=TC.
\end{align}
The space of the matrices $A$ is 
$R^1_1$ for $s_T=+$, $s_C=+$; 
$R^0_2$ for $s_T=+$, $s_C=-$; 
$R^1_1$ for $s_T=-$, $s_C=+$; and
$R^2_0$ for $s_T=-$, $s_C=-$.
For $s_{TC}=+$,
$C$ will commute with both $A$ and $T$.
We find  space of the matrices $A$ to be
$R^0_1$ for $s_T=+$, $s_C=+$; 
$R^1_0$ for $s_T=-$, $s_C=+$; and
$C_1$ for $s_T=\pm$, $s_C=-$.

If there are $U(1)$, time reversal, and
charge conjugation symmetries
(full symmetry group =
 $G_{s_Ts_C}^{s_{UT}s_{TC}}(U,T,C)$), then for
$s_{TC}=s_{UT}=-$, $A$ satisfies
\begin{align}
&
 A^2=-1, \    A\rho_i+\rho_i A=0,\ 
\rho_1^2= \rho_2^2=s_T, \ 
\rho_3^2=-s_Ts_C, 
\nonumber\\
&
\rho_1=T, \ \ \rho_2=TQ,
 \ \ \rho_3=TC.
\end{align}
The space of the matrices $A$ is 
$R^1_2$ for $s_T=+$, $s_C=+$; 
$R^0_3$ for $s_T=+$, $s_C=-$; 
$R^2_1$ for $s_T=-$, $s_C=+$; and
$R^3_0$ for $s_T=-$, $s_C=-$.

For $s_{UT}=-, \ s_{TC}=+$, 
$A$ satisfies
\begin{align}
&
 A^2=-1, \   A\rho_i+\rho_i A=0,\  
\rho_1^2= \rho_2^2=s_T,  \ 
\rho_3^2=-s_Ts_C, 
\nonumber\\
&
\rho_1=T, \ \ \rho_2=TQ,
 \ \ \rho_3=TQC.
\end{align}
The space of the matrices $A$ is
$R^1_2$ for $s_T=+$, $s_C=+$; 
$R^0_3$ for $s_T=+$, $s_C=-$; 
$R^2_1$ for $s_T=-$, $s_C=+$; and
$R^3_0$ for $s_T=-$, $s_C=-$.

For $s_{UT}=+, \ s_{TC}=-$,
$A$ satisfies
\begin{align}
&
 A^2=-1, \    A\rho_i+\rho_i A=0,\   
\rho_1^2=s_T, \  
\rho_2^2= \rho_3^2=-s_Ts_C, 
\nonumber\\
&
\rho_1=T, \  \rho_2=TC,
 \  \rho_3=TCQ.
\end{align}
The space of the matrices $A$ is
$R^2_1$ for $s_T=+$, $s_C=+$; 
$R^0_3$ for $s_T=+$, $s_C=-$; 
$R^1_2$ for $s_T=-$, $s_C=+$; and
$R^3_0$ for $s_T=-$, $s_C=-$.

For $s_{UT}=+,\ s_{TC}=+$, we find that $A$ satisfy
\begin{align}
&
A^2=-1, \   
A\rho_i+\rho_i A=0,\ 
\rho_1^2=-s_T, \ 
\rho_2^2= \rho_3^2=s_Ts_C, 
\nonumber\\
&
\rho_1=TQ,\   
\rho_2=TC,\ 
\rho_3=TCQ
.
\end{align}
We see that the matrices $A$
form a space 
$R^1_2$ for $s_T=+,\ s_C=+$;
$R^3_0$ for $s_T=+,\ s_C=-$;
$R^2_1$ for $s_T=-,\ s_C=+$; and
$R^0_3$ for $s_T=-,\ s_C=-$;

\subsection{The properties of classifying spaces}

In general, we can consider a real antisymmetric matrix $A$ that satisfies
(for fixed real matrices $\rho_i$, $i=1,...,p+q$)
\begin{align}
 &
A=\rho_{p+q+1}, \ \ \ \ \rho_j\rho_i+\rho_i \rho_j=\Big|_{i\neq j}\ 0,
\nonumber\\
&
\rho_i^2=\Big|_{i=1,...,p}\ 1, \ \ \
\rho_i^2=\Big|_{i=p+1,...,p+q+1}\ -1.
\end{align}
The space of those $A$ matrices is denoted as $R^q_p$.  

Let us show that
\begin{align}
\label{pqpq11}
 R^q_p=R^{q+1}_{p+1} .
\end{align}
From $\t A \in R^q_p$ that satisfies
the following Clifford algebra $Cl(p,q+1)$
\begin{align}
&
\label{pq}
 \t A=\t\rho_{p+q+1}, \ \ \t\rho_j\t\rho_i+\t\rho_i \t\rho_j=\Big|_{i\neq j}\ 0,
\nonumber\\
&
\t\rho_i^2=\Big|_{i=1,...,p}\ 1, \ \
\t\rho_i^2=\Big|_{i=p+1,...,p+q+1}\ -1,
\end{align}
we can define
\begin{align}
&
 \rho_i=\Big|_{i=1,...,p}\ \t\rho_i\otimes \si^z, \ \ 
 \rho_{p+1}= I\otimes \si^x, 
\nonumber\\
&
 \rho_i=\Big|_{i=p+2,...,p+q+2}\ \t\rho_{i-1} \otimes \si^z, \ \  
 \rho_{p+q+3}= I\otimes \veps.
\end{align}
We can check that such $\rho_i$ satisfy
the following Clifford algebra $Cl(p+1,q+2)$
\begin{align}
\label{pq11}
&
\rho_j\rho_i+\rho_i \rho_j=\Big|_{i\neq j}\ 0,
\nonumber\\
&
\rho_i^2=\Big|_{i=1,...,p+1}\ 1, \ \ 
\rho_i^2=\Big|_{i=p+2,...,p+q+3}\ -1.
\end{align}
If we fix $\rho_i$, $i\neq p+q+2$, then the space formed by $A=\rho_{p+q+2}$
satisfying the above condition is given by $R^{q+1}_{p+1}$.  The above
construction gives rise to a map from $R^q_p \to R^{q+1}_{p+1}$.  Since
$A=\rho_{p+q+2}$ satisfying \eqn{pq11} must has a form $\t A\otimes \si^z$, with
$\t A$ satisfying \eqn{pq}.  This gives us a map $R^{q+1}_{p+1} \to R^q_p$.
Thus $R^{q+1}_{p+1} = R^q_p$.

We can also consider real
symmetric matrix $A$ that satisfies (for fixed real matrices $\rho_i$,
$i=1,...,p$)
\begin{align}
 A=\rho_{p+1}, \ \ \ \ \rho_j\rho_i+\rho_i \rho_j=\Big|_{i\neq j}\ 0,\ \ \
\rho_i^2=\Big|_{i=1,...,p+1}\ 1 .
\end{align}
The space of those matrices is denoted as $R_p$.  

\begin{table*}[tb]
 \centering
 \begin{tabular}{ |c||c|c|c|c|c|c|c|c| }
 \hline
$p$ mod 8 &  $0$ & $1$ & $2$ & $3$ & $4$ & $5$ & $6$ & $7$   \\ 
 \hline
$R_p$ &
 $\frac{O(l+m)}{O(l)\times O(m)}\times \Z$ & 
 $O(n)$ & 
 $\frac{O(2n)}{U(n)}$ & 
 $\frac{U(2n)}{Sp(n)}$ & 
 $\frac{Sp(l+m)}{Sp(l)\times Sp(m)}\times \Z$ & 
 $Sp(n)$ & 
 $\frac{Sp(n)}{U(n)}$ & 
 $\frac{U(n)}{O(n)}$  \\ 
\hline
$\pi_0(R_p)$ &
$\Z$ &
$\Z_2$ &
$\Z_2$ &
$0$ &
$\Z$ &
$0$ &
$0$ &
$0$ \\ 
$\pi_1(R_p)$ &
$\Z_2$ &
$\Z_2$ &
$0$ &
$\Z$ &
$0$ &
$0$ &
$0$ & 
$\Z$ \\
$\pi_2(R_p)$ &
$\Z_2$ &
$0$ &
$\Z$ &
$0$ &
$0$ &
$0$ & 
$\Z$ &
$\Z_2$ \\
$\pi_3(R_p)$ &
$0$ &
$\Z$ &
$0$ &
$0$ &
$0$ & 
$\Z$ &
$\Z_2$ &
$\Z_2$ \\
$\pi_4(R_p)$ &
$\Z$ &
$0$ &
$0$ &
$0$ &
$\Z$ &
$\Z_2$ &
$\Z_2$ &
$0$ \\
$\pi_5(R_p)$ &
$0$ &
$0$ &
$0$ &
$\Z$ &
$\Z_2$ &
$\Z_2$ &
$0$ &
$\Z$ \\
$\pi_6(R_p)$ &
$0$ &
$0$ &
$\Z$ &
$\Z_2$ &
$\Z_2$ &
$0$ &
$\Z$ &
$0$ \\
$\pi_7(R_p)$ &
$0$ &
$\Z$ &
$\Z_2$ &
$\Z_2$ &
$0$ &
$\Z$ &
$0$ &
$0$ \\
\hline
 \end{tabular}
 \caption{
The spaces $R_p$ and their homotopy groups $\pi_d(R_p)$.
}
 \label{Rppi0}
\end{table*}

In the following, we are
going to show that
\begin{align}
\label{qq2}
 R^q_0=R_{q+2} .
\end{align}
From $\t A \in R^q_0$ that satisfies
the Clifford algebra $Cl(0,q+1)$
\begin{align}
\label{q1}
 \t A=\t\rho_{q+1}, \ \ \ \ \t\rho_j\t\rho_i+\t\rho_i \t\rho_j=\Big|_{i\neq j}\ 0,\ \ \
\t\rho_i^2=\Big|_{i=1,...,q+1}\ -1,
\end{align}
we can define
\begin{align}
 \rho_i=\Big|_{i=1,...,q+1}\ \t\rho_i \otimes \veps, \ \ \ \
 \rho_{q+2}= I\otimes \si^z, \ \ \ \
 \rho_{q+3}= I\otimes \si^x.
\end{align}
we can check that
$\rho_i$ form the Clifford algebra $Cl(q+3,0)$
\begin{align}
\label{q2}
 \rho_j\rho_i+\rho_i \rho_j=\Big|_{i\neq j}\ 0,\ \ \
\rho_i^2=\Big|_{i=1,...,q+3}\ 1 .
\end{align}
If we fix $\rho_i$, $i\neq q+1$, then the space 
formed by $A=\rho_{q+1}$ satisfying
the above condition is given by
$R_{q+2}$.  The above construction gives rise to a map from $R^q_0
\to R_{q+2}$.  Since $A=\rho_{q+1}$ satisfying \eqn{q2} must has a form $\t
A\otimes \veps$, with $\t A$ satisfying \eqn{q1}.  This gives us a map
$R_{q+2} \to R^q_0$.  Thus $R^q_0=R_{q+2}$.

In addition we also have the following periodic relations
\begin{align}
\label{pq8}
 R^q_p= 
 R^{q+8}_p= 
 R^q_{p+8},\ \ \ \ \ R_p=R_{p+8}. 
\end{align}
This can be shown by noticing the following
16 dimensional real symmetric representation of Clifford algebra $Cl(0,8)$:
\begin{align}
  \th_1&=\veps\otimes\si^z\otimes \si^0\otimes \veps,
& 
  \th_2&=\veps\otimes\si^z\otimes \veps \otimes \si^x,
\nonumber\\
  \th_3&=\veps\otimes\si^z\otimes \veps \otimes \si^z, 
& 
   \th_4&=\veps\otimes\si^x\otimes \veps \otimes \si^0,
\nonumber\\
   \th_5&=\veps\otimes\si^x\otimes \si^x \otimes \veps,
&
   \th_6&=\veps\otimes\si^x\otimes \si^z \otimes \veps,
\nonumber\\
   \th_7&=\veps\otimes\veps \otimes \si^0 \otimes \si^0,
&
   \th_8&=\si^x\otimes\si^0 \otimes \si^0 \otimes \si^0,
\end{align}
which satisfy
\begin{align}
 \th_i\th_j + \th_j\th_i =\Big|_{i\neq j}\ 0, \ \ \ \ \ \ \ \ \
 \th_i^2 =\Big|_{i=0,...,8}\ 1.
\end{align}
We find that $\th= \th_1 \th_2 \th_3 \th_4 \th_5 \th_6 \th_7 \th_8
=\si^z\otimes\si^0 \otimes \si^0 \otimes \si^0 $ anticommute with $\th_i$.
From $\t A \in R^q_p$ that satisfies \eqn{pq},
we can define
\begin{align}
&
 \rho_i=\Big|_{i=1,...,p}\ \t\rho_i\otimes \th, \ \ \ 
 \rho_{p+i}=\Big|_{i=1,...,8} I\otimes \th_i, 
\nonumber\\
&
 \rho_i=\Big|_{i=p+9,...,p+q+9}\ \t\rho_{i-8} \otimes \si^z.
\end{align}
We can check that such $\rho_i$ satisfy
\begin{align}
\label{p8q}
&
\rho_j\rho_i+\rho_i \rho_j=\Big|_{i\neq j} 0,\ \ \
\rho_i^2=\Big|_{i=1,...,p+8} 1, 
\nonumber\\
&
\rho_i^2=\Big|_{i=p+9,...,p+q+9} -1.
\end{align}
If we fix $\rho_i$, $i\neq p+q+9$, then the space formed by $A=\rho_{p+q+9}$
satisfying the above condition is given by $R^{q}_{p+8}$.  The above
construction gives rise to a map from $R^q_p \to R^q_{p+8}$.  On the other
hand, the matrix that anticommute with all $\th_i$'s must be proportional to
$\th$. Thus $A=\rho_{p+q+9}$ satisfying \eqn{p8q} must has a form $\t A\otimes
\th$, with $\t A$ satisfying \eqn{pq}.  This gives us a map $R^q_{p+8} \to
R^q_p$.  Thus $R^q_{p+8} = R^q_p$.  Using a similar approach, we can show
$R_p=R_{p+8}$.  The relation \eqn{pqpq11}, \eqn{qq2}, and \eqn{pq8} allow us
show
\begin{align}
 R^q_p=R_{q-p+2 \text{ mod }8}.
\end{align}
So we can study the space $R^q_p$ via the space $R_{q-p+2 \text{ mod }8}$.

Let us construct some of the $R_p$ spaces.  $R_0$ is formed by real symmetric
matrices $A$ that satisfy $A^2=1$. Thus $A$ has a form $O \bpm I_{l\times l} &
0\\ 0 & -I_{m\times m} \epm O^{-1}$, $O\in O(l+m)$.  We see that $R_0=\bigcup_m
O(l+m)/O(l)\times O(m)= \frac{O(l+m)}{O(l)\times O(m)}\times \Z$.  

$R_1$ is
formed by real symmetric matrices $A$ that satisfy $A^2=1$ and $A\rho_1=-\rho_1
A$ with $\rho_1=\si^z\otimes I_{n\times n}$.
Thus $A$ has a form 
\begin{align}
A&=
\e^{\si^z\otimes M_{n\times n}}
\e^{\si^0\otimes L_{n\times n}}
[\si^x\otimes I_{n\times n}]
\e^{-\si^0\otimes L_{n\times n}}
\e^{-\si^z\otimes M_{n\times n}}
\nonumber\\
&=
\e^{\si^z\otimes M_{n\times n}}
[\si^x\otimes I_{n\times n}]
\e^{-\si^z\otimes M_{n\times n}}
,
\end{align}
where $ \e^{\si^0\otimes L_{n\times n}} \in O(n)$ and $ \e^{\si^z\otimes
M_{n\times n}} \in O(n)$ are the transformations that leave $\rho_1$ unchanged.
We see that $R_1=O(n)$.  The other spaces $R_p$ and $\pi_0(R_p)$
are listed in Table \ref{Rppi0}. 
Note that for space $S\times \Z$, we have $\pi_0(S\times
Z)=\pi_0(S)\times \Z$. Also $O(n)$ in dividend usually leads to the $\Z_2$ in
$\pi_0$. Otherwise $\pi_0=\{ 0\}$.  $O(n)$ in dividend can give rise to $\Z_2$
because  $O\in O(n)$ with $\det(O)=1$ and $\det(O)=-1$ cannot be smoothly
connected. $O(l+m)$ in $ \frac{O(l+m)}{O(l)\times O(m)}$ does not lead to
$\Z_2$ because for $O\in O(l+m)$, we can change the sign of $\det(O)$ by
multiplying $O$ with an element in $O(l)$ [or $O(m)$].

For free fermion systems in $0$-dimension with no symmetry and no fermion
number conservation, the classifying space is $R^0_0$.  Since
$\pi_0(R^0_0)=\pi_0(R_2)=\Z_2$, such free fermion systems has two possible
gapped phases.  One phase has even numbers of fermions in the ground state and
the other phase has odd numbers of fermions in the ground state. (Note that the
fermion number mod 2 is still conserved even without any symmetry.)

For free electron systems in $0$-dimension with time reversal symmetry and
electron number conservation (the symmetry group $G_-^-(U,T)$), the
classifying space is $R^2_0$.  Since $\pi_0(R^2_0)=\pi_0(R_4)=\Z$, the possible
gapped phases are labeled by an integer $n$.  The ground state has $2n$
fermions.  The electron number in the ground state is always even due to the
Kramer degeneracy.  

If we drop the electron number conservation (the symmetry group becomes
$G_-(T)$), then the ground state will have uncertain but even numbers of
electrons.  The ground state cannot have odd numbers of electrons.  This
implies that free electron systems with only time reversal symmetry in
$0$-dimension has only one possible gapped phase.  This agrees with
$\pi_0(R^1_0)=\pi_0(R_3)=\{ 0\}$, where $R^1_0$ is the classifying space for
symmetry group $G_-(T)$.

\subsection{The $d\neq 0$ cases}

Now let us consider the $d\neq 0$ cases.  Again, let us first assume that the
fermion system described by $\hat H=\frac{\imth}{4} \sum_{IJ} A_{IJ}\hat
\eta_I\hat \eta_J$ has the translation symmetry, as well as the time reversal
symmetry and fermion number conservation.  We also assume that the single-body
energy bands of antisymmetric hermitian matrix $\imth A$ have some Dirac points
at zero energy and there are no other zero energy states in the Brillouin zone.
The gapless single-body excitations in the system are described by the
continuum limit of $\imth A$:
\begin{align}
 \imth A = \imth \sum_{i=1}^d \ga_i \prt_i ,
\end{align}
where we have folded all the Dirac points to the $\v k=0$ point.  Without
losing generality, we have also assumed that all the Dirac points have the same
velocity.  
Since $\prt_i$ is real and antisymmetric, thus
$\ga_i$, $i=1,...,d$ are real symmetric $\ga$-matrices (of
infinity dimension) that satisfy:
\begin{align}
\ga_i\ga_j+\ga_j\ga_i=2\del_{ij} ,\ \ \ \ 
\ga_i^*=\ga_i.
\end{align}
Again, the allowed $\ga_i$ always satisfy the condition that $M^2=-1$ and
$\ga_iM+M\ga_i=0$ has a solution for $M$.  Since the time reversal and the
$U(1)$ transformations do not affect $\prt_i$, therefore the symmetry
conditions on $A$, $ A T + TA =0$ and $ A Q - QA =0$, become the symmetry
conditions on the $\ga$-matrices:
\begin{align}
 \ga_i T + T\ga_i =0,\ \ \ \ 
\ga_i Q - Q\ga_i =0
.
\end{align}

Now we add perturbations that may break the translation, time reversal and the $U(1)$ symmetries, and ask: how
many different ways are there to gap the Dirac points.  The Dirac points can be
fully gapped by real antisymmetric mass matrices $M$ that satisfy
\begin{align}
\label{Mgai}
 \ga_i M + M\ga_i=0,
\end{align}
The resulting single-body Hamiltonian becomes
$ \imth A = \imth \sum_{i=1}^d [\ga_i \prt_i +M ]$.

If there is no symmetry, we only require the real antisymmetric mass matrix $M$
to be invertible (in addition to \eqn{Mgai}). Without losing generality, we can
choose the mass matrix to also satisfy 
\begin{align}
 M^2=-1.
\end{align}
The space of those mass matrices is given by $R^0_d$.

If there are some symmetries, the real antisymmetric mass matrix $M$
also satisfy some additional condition, as discussed before:
$M$ anticommutes with a set of $p+q$ matrices $\rho_i$ that anticommute
among themselves with $p$ of them square to 1 and $q$ of them square to -1.
The number of $p,q$ depend on full symmetry group $G_f$.
Since $\ga_i$ do not break the symmetry, so, just like $M$,
$\ga_i$ also anticommute $\rho_i$.
So, in total, $M$ anticommutes with a set of $p+q+d$ matrices $\rho_i$
and $\ga_i$ that anticommute
among themselves with $p+d$ of them square to 1 and $q$ of them square to -1.
Those mass matrices from a space $R^q_{p+d}$.

The different disconnected components of $R^q_{p+d}$ represent different
``bulk'' gapped phases of the free fermions.  Thus, the ``bulk'' gapped phases
of the free fermions in $d$-dimensions are classified by
$\pi_0(R^q_{p+d})=\pi_0(R_{q-p-d+2 \text{ mod }8})$, with $(p,q)$ depending on
the symmetry. The results are summarized in Table \ref{GFFsym}.  

\subsection{A general discussion}
\label{gd}

Now, let us give a gneral discussion of the classifying problem of free fermion
systems.  To classify the gapped phases of free fermion Hamiltonian
we need to construct the space of antisymmetric mass matrix $M$
that satisfy
\begin{align}
 M^2=-1 .
\end{align}
The  mass matrices $A$ always anti commute with $\ga$-matrices $\ga_i$,
$i=1,2,...,d$.
When the mass matrices $M$ has some symmeties, then the mass matrices satisfy
more linear conditions.  Let us assume that all those conditions can be
expressed in the following form
\begin{align}
 M\rho_i=-\rho_i M,\ \ \ \ \rho_i \rho_j=-\rho_j\rho_i,
\nonumber\\
MU_I=U_I M,\ \ \ \ U_I \rho_i=\rho_iU_I,
\end{align}
where $\rho_i$ and $U_I$ are real matrices labeled by $i$ and $I$, and
$\ga_1,...,\ga_d$ are included in $\rho_i$'s.  If we have another symmetry
condition $W$ such that $ MW=-W M$ and $W\rho_{i_0}=\rho_{i_0} W$ for a
partcular $i_0$, Then $U=W\rho_{i_0}$ will commute with $M$ and $\rho_i$, and
will be part of $U_I$.

$U_I$ will form some algebra. Let us use $\al$ to label the irriducible
representations of the algebra.  Then the one fermion Hilbert space has a form
$\cH =\oplus_\al \cH_\al\otimes \cH_\al^0$, where the space $\cH_\al^0$ forms
the $\al^\text{th}$ irreducible representations of the algebra.  For such a
decomposition of the Hilbert space, $M$ has the following block diagonal form
\begin{align}
 M=\oplus_\al (M^\al\otimes I^\al) ,
\end{align}
where $I^\al$ acts within $\cH_\al^0$ as an identity operator, and $M^\al$ acts
within $\cH_\al$.  $\rho_i$'s have a similar form
\begin{align}
 \rho_i=\oplus_\al (\rho_i^\al\otimes I_\al) ,
\end{align}
where $\rho_i^\al$ act
within $\cH_\al$.
So within the Hilbert space $\cH_\al$, we have
\begin{align}
\label{Malrhoal}
 M^\al\rho_i^\al=-\rho_i^\al M^\al,\ \ \ \ \rho_i^\al \rho_j^\al=-\rho_j^\al\rho_i^\al.
\end{align}
What we are trying to do in this paper is actually to contruct the space of
$M^\al$ matrices that satisfy the condition \eqn{Malrhoal}.

If fermions only form one irreducible representation of the $U_I$ algebra,
then the classifying space of $M_\al$ and $M$ will be the same.  The results of
this paper (such as Tables \ref{GFF}, \ref{GFFsym}) are obtained under such an
assumption.

If fermions form $n$ distinct irreducible representations of the $U_I$ algebra,
then the classifying space of $M$ will be $R^n$, where $R$ is the classifying
of $M^\al$ constructed in this paper.  Note that the classifying spaces of
$M^\al$ are the same for different irreducible representations and hence $R$ is
independent of $\al$.  So if $M_\al$'s are classified by $\Z_k$, $k=1,2$, or
$\infty$, then $M$'s are  classified by $\Z_k^n$.

To illustrate the above result,
let us use the symmetry $G_+(C)=Z_2^C\times Z_2^f$
as an example. If the fermions
form one irreducible representation of $Z_2^C$, for example,
$\hat C c_i \hat C^{-1}=-c_i$,
then the non-interacting symmetric gapped phases
are classified by
\begin{align}
 \bmm d:& 0 & 1 & 2 & 3 & 4 &  5 & 6& 7\\
  \text{gapped phases}: &\Z_2 & \Z_2 & \Z & 0 & 0 &  0 & \Z& 0  \\
\emm
\end{align}
which is the result in Table \ref{GFFsym}.
If the fermions form both the irreducible representations of $Z_2^C$, $\hat C
c_{i+} \hat C^{-1}=+c_{i+}$ and $\hat C c_{i-} \hat C^{-1}=-c_{i-}$ (\ie one
type of fermions carries $Z_2^C$-charge 0 and another type of fermions carries
$Z_2^C$-charge 1), then the non-interacting symmetric gapped phases are
classified by
\begin{align}
 \bmm d:& 0 & 1 & 2 & 3 & 4 &  5 & 6& 7\\
  \text{gapped phases}: &\Z_2^2 & \Z_2^2 & \Z^2 & 0 & 0 &  0 & \Z^2 & 0  \\
\emm
\end{align}
The four $d=0$ phases correspond to the ground state with even or odd
$Z_2$-charge-0 fermions and even or odd $Z_2$-charge-1 fermions.  The four
$d=1$ phases correspond to the phases where the $Z_2$-charge-0 fermions are in
the trvial or non-trivial phases of Majorana chain and the $Z_2$-charge-1
fermions are in the trvial or non-trivial phases of Majorana chain.  The $d=2$
phase labeled by two integers $(m,n)\in \Z^2$ corresponds to the phase where
the $Z_2$-charge-0 fermions have $m$ right moving Majorana chiral modes and the
$Z_2$-charge-1 fermions have $n$ right moving Majorana chiral modes.  (if $m$
and/or $n$ are negative, we then have the corresponding number of left moving
Majorana chiral modes.)

Some of the above gapped phases have intrinsic fermionic topological orders.
So only a subset of them are non-interacting fermionic SPT phases:
\begin{align}
 \bmm d_{sp}:& 0 & 1 & 2 & 3 & 4 &  5 & 6& 7\\
  \text{SPT phases}: &\Z_2 & \Z_2 & \Z & 0 & 0 &  0 & \Z & 0  \\
\emm
\end{align}
The two $d_{sp}=0$ phases correspond to the ground states with even numbers of
fermions and 0 or 1 $Z_2$-charges.  The two $d_{sp}=1$ phases correspond to the
phases where the $Z_2$-charge-0 fermions and the $Z_2$-charge-1 fermions are
both in the trvial or non-trivial phases of Majorana chain.  The $d_{sp}=2$
phase labeled by one integers $n\in \Z$ corresponds to the phase where the
$Z_2$-charge-0 fermions have $n$ right moving Majorana chiral modes and the
$Z_2$-charge-1 fermions have $n$ left moving Majorana chiral modes.

\section{Classification with translation symmetry}

\begin{table*}[t]
 \centering
 \begin{tabular}{ |c||c|c||c|c||c| }
 \hline
Symmetry & $C_{p_G}$ & $p_G$ &  
$\bmm
\text{point}\\ 
\text{defect} 
\emm$
& 
$\bmm
\text{line}\\ 
\text{defect} 
\emm$
&
example phases
\\
\hline
\hline
{
\footnotesize
$\bmm
\red{U(1)}\\[1mm]
G_-(C)
\emm$
} & $\frac{U(l+m)}{U(l)\times U(m)}\times \Z$ & 0 
& $0$ & $\Z$ & 
$\bmm
\text{(Chern)}\\
\text{insulator}
\emm$
~~
$\bmm
\text{supercond.}\\[-1mm]
\text{with collinear}\\[-1mm]
\text{spin order}
\emm$
\\
 \hline
{
\footnotesize
$\bmm
\red{G_\pm^+(U,T)}\\[1mm]
G_{--}^+(T,C)\\[1mm]
G_{+-}^+(T,C)
\emm$
}  & $U(n)$ & 1
& $\Z$ & $0$ & 
$\bmm
\text{supercond. w/ real pairing}\\[-1mm]
\text{and $S_z$ conserving}\\[-1mm]
\text{spin-orbital coupling}
\emm$
\\
\hline
 \end{tabular}
\caption{
(Color online) Classification of point defects and line defects that have some
symmetries in gapped phases of non-interacting fermions.
``0'' means that there is no non-trivial topological defects.  $\Z$ means that
topological non-trivial defects plus the topological trivial defect are labeled
by the elements in $\Z$.
Non-trivial topological defects have protected gapless excitations,
while trivial topological defects have no protected gapless excitations.
}
 \label{defC}
\end{table*}

\begin{table*}[tb]
{
 \begin{tabular}{ |c||c|c|c|c|c|c|c|c| }
 \hline
Symm.
&
{\scriptsize
$\bmm
\red{G_{+}^{-}(U,T)}\\[1mm]
G_{+-}^{-}(T,C)
\emm$
}%
& 
{\scriptsize
$\bmm
\red{G_+(T)}\\[1mm]
G_{++}^{+}(T,C)\\[1mm]
\red{G_{++}^{--}(U,T,C)}\\[1mm]
G_{++}^{-+}(U,T,C)\\[1mm]
G_{-+}^{+-}(U,T,C)\\[1mm]
G_{++}^{++}(U,T,C)
\emm$
}%
&
{\scriptsize
$\bmm
\text{{\small\red{``none''}}}\\[1mm]
G_+(C)\\[1mm]
G_{++}^{-}(T,C)\\[1mm]
G_{-+}^{-}(T,C)\\[1mm]
G_+(C)\\[1mm]
G_+(U,C)
\emm$
}%
&
{\scriptsize
$\bmm
\red{G_-(T)}\\[1mm]
G_{-+}^{+}(T,C)\\[1mm]
G_{-+}^{--}(U,T,C)\\[1mm]
G_{-+}^{-+}(U,T,C)\\[1mm]
G_{++}^{+-}(U,T,C)\\[1mm]
G_{-+}^{++}(U,T,C)
\emm$
}%
&
{\scriptsize
$\bmm
\red{G_{-}^{-}(U,T)}\\[1mm]
G_{--}^{-}(T,C)
\emm$
}%
&
{\scriptsize
$\bmm
G_{--}^{--}(U,T,C)\\[1mm]
\red{G_{--}^{-+}(U,T,C)}\\[1mm]
G_{--}^{+-}(U,T,C)\\[1mm]
G_{+-}^{++}(U,T,C)
\emm$
}%
&
{\scriptsize
$\bmm
\red{G_{-}(U,C)}\\[1mm]
\red{SU(2)}
\emm$
}
&
{\scriptsize
$\bmm
G_{+-}^{--}(U,T,C)\\[1mm]
G_{+-}^{-+}(U,T,C)\\[1mm]
G_{+-}^{+-}(U,T,C)\\[1mm]
\red{G_{--}^{++}(U,T,C)}\\[1mm]
\red{G[SU(2),T]}
\emm$
}%
\\
 \hline
$R_{p_G}$ &
{\footnotesize
$\frac{O(l+m)}{O(l)\times O(m)}\times \Z$
}
& 
 $O(n)$ & 
 $\frac{O(2n)}{U(n)}$ & 
 $\frac{U(2n)}{Sp(n)}$ &
{\footnotesize
$\frac{Sp(l+m)}{Sp(l)\times Sp(m)}\times \Z$
}
& 
 $Sp(n)$ & 
 $\frac{Sp(n)}{U(n)}$ & 
 $\frac{U(n)}{O(n)}$  \\ 
 \hline
$p_G$: &  $0$ & $1$ & $2$ & $3$ & $4$ & $5$ & $6$ & $7$   \\ 
\hline
$\bmm
\text{point}\\
\text{defect}
\emm$ & 
$0$ & $\Z$ & $\Z_2$ & $\Z_2$ & $0$ & $\Z$ & $0$ & $0$ \\
\hline
$\bmm
\text{line}\\
\text{defect}
\emm$ & 
$0$ & $0$ & $\Z$ & $\Z_2$ & $\Z_2$ & $0$ & $\Z$ & $0$ \\
\hline
$\bmm
\text{example}\\
\text{phases}
\emm$ 
&
{\footnotesize
$\bmm
\text{insulator}\\
\text{w/ coplanar}\\
\text{spin order}
\emm$
}
&
{\footnotesize
$\bmm
\text{supercond.}\\
\text{w/ coplanar}\\
\text{spin order}
\emm$
}
&
{\footnotesize
$\bmm
\text{supercond.}
\emm$
}
&
{\footnotesize
$\bmm
\text{supercond.}\\
\text{w/ time}\\
\text{reversal}\\
\emm$
}
&
{\footnotesize
$\bmm
\text{insulator}\\
\text{w/ time}\\
\text{reversal}\\
\emm$
}
&
{\footnotesize
$\bmm
\text{insulator}\\[-1mm]
\text{w/ time}\\[-1mm]
\text{reversal and}\\[-1mm]
\text{inter-sublattice}\\[-1mm]
\text{hopping}
\emm$
}
&
{\footnotesize
$\bmm
\text{spin}\\
\text{singlet}\\
\text{supercond.}\\
\emm$
}
&
{\footnotesize
$\bmm
\text{spin}\\[-1mm]
\text{singlet}\\[-1mm]
\text{supercond.}\\[-1mm]
\text{w/ time}\\[-1mm]
\text{reversal}
\emm$
}
\\
\hline
\end{tabular}
}
 \caption{(Color online) Classification of point defects and line defects that have some
symmetries in gapped phases of non-interacting fermions.   ``0'' means that
there is no non-trivial topological defects.  $\Z_n$ means that topological
non-trivial defects plus the topological trivial defect are labeled by the
elements in $\Z_n$.
Non-trivial defects have protected gapless excitations in them.
}
 \label{defR}
\end{table*}

For the free fermion systems with certain internal symmetry $G_f$, we have
shown that their gapped phases are classified by $\pi_0(R_{p_G-d})$ or
$\pi_0(C_{p_G-d})$ in $d$ dimensions, where the value of $p_G$ is determined
from the full symmetry group $G_f$.  We know that $\pi_0$ is an Abelian group
with commuting  group multiplication ``$+$'': $a,b \in \pi_0$ implies that $a+b
\in \pi_0$.  The ``$+$'' operation has a physical meaning.  If two ``bulk''
gapped fermion systems are labeled by $a$ and $b$ in $\pi_0$, then stacking the
two systems on together will give us a new ``bulk'' gapped fermion system
labeled by $a+b \in \pi_0$.

We note that the classification by $\pi_0(R_{p_G-d})$ or $\pi_0(C_{p_G-d})$ is
obtained by assume there is no translation symmetry.  In the presence of
translation symmetry, the gapped phases are classified
differently.\cite{R0922,FKM0703,MB0706,R1054} However, the new classification
can be obtained from $\pi_0(R_{p_G-d})$. For the free fermion systems with
internal symmetry $G_f$ and translation symmetry, their gapped phases are
classified by\cite{K0886}
\begin{align}
 \prod_{k=0}^d [\pi_0(R_{p_G-d+k})]^{\binom{d}{k}} ,
\end{align}
where $\binom{d}{k}$ is the binomial coefficient.
The above is for the real classes. For the complex classes, we have a similar 
classification:
\begin{align}
 \prod_{k=0}^d [\pi_0(C_{p_G-d+k})]^{\binom{d}{k}} .
\end{align}

Such a result is obtained by stacking the lower dimensional topological phases
to obtain higher dimensional ones.  For 1-dimensional free fermion systems with
internal symmetry $G_f$, their gapped phases are classified by $\pi_0(R_{p_G-1})$.
We can also have a 0-dimensional gapped phase on each unit cell of the
1-dimensional system if there is a translation symmetry.  The 0-dimensional
gapped phases are classified by $\pi_0(R_{p_G})$. Thus the combined gapped
phases (with translation symmetry) are classified by $\pi_0(R_{p_G-1})\times
\pi_0(R_{p_G})$.  In 2-dimensions, the gapped phases are classified by
$\pi_0(R_{p_G-2})$.  The gapped phases on each unit cell are classified by
$\pi_0(R_{p_G})$.  Now we can also have 1-dimensional gapped phases on the
lines in $x$-direction, which are classified by $\pi_0(R_{p_G-1})$. The same
thing for the lines in $y$-direction. So the combined gapped phases (with
translation symmetry) are classified by
$\pi_0(R_{p_G-2})\times[\pi_0(R_{p_G-1})]^2\times \pi_0(R_{p_G})$.  In three
dimensions, the translation symmetric gapped phases are classified by
$\pi_0(R_{p_G-3})\times[\pi_0(R_{p_G-2})]^3\times[\pi_0(R_{p_G-1})]^3\times
\pi_0(R_{p_G})$.

\section{Defects in $d$-dimensional gapped free fermion phases
with symmetry $G_f$}

For the $d$-dimensional free fermion systems with internal symmetry $G_f$, we
have shown that their ``bulk'' gapped Hamiltonians (or the mass matrices) form
a space $R_{p_G-d}$ or $C_{p_G-d}$.  (More precisely, the gapped Hamiltonians
or the mass matrices form a space that is homotopically equivalent to
$R_{p_G-d}$ or $C_{p_G-d}$). From the space $R_{p_G-d}$ or $C_{p_G-d}$, we find
that the point defects that have the symmetry $G_f$ are classified by
$\pi_{d-1}(R_{p_G-d})$ or $\pi_{d-1}(C_{p_G-d})$.

Physically, there is another way to classify point defects: we can simply add a
segment of 1D ``bulk'' gapped free fermion hopping system with the same
symmetry to the $d$-dimensional system.  Since the translation symmetry is not
required, the new $d$-dimensional system still belong to the same symmetry
class. There are finite bulk gap away from the two ends of the added 1D
segment.  So the new $d$-dimensional system may contain two non-trivial
defects.  The classes of defect is classified by the classes of the added 1D
``bulk'' gapped free fermion hopping system.  So we find that the point defects
that have the symmetry $G_f$ are also classified by $\pi_{0}(R_{p_G-1})$ or
$\pi_0(C_{p_G-1})$. 

Similarly, the line defects that have the symmetry $G_f$ are classified by
$\pi_{d-2}(R_{p_G-d}) $ or $\pi_{d-2}(C_{p_G-d})$.  
Again, we can also create line defects by adding
a disk of 2D ``bulk'' gapped free fermion hopping system to the
original $d$-dimensional system.
This way, we find that the
line defects that have the symmetry $G_f$ are also classified by
$\pi_{0}(R_{p_G-2})$ or $\pi_0(C_{p_G-2})$. 

In general, the defects with dimension $d_0$ are classified by
$\pi_{d-d_0-1}(R_{p_G-d})$ or $\pi_{d-d_0-1}(C_{p_G-d})$, or equivalently by 
$\pi_{0}(R_{p_G-d_0-1})$ or $\pi_0(C_{p_G-d_0-1})$. 
In order for the above physical picture to be consistent, we require that
\begin{align}
 \pi_{n}(R_{p_G})= \pi_{0}(R_{p_G+n}), \ \ \ \
 \pi_{n}(C_{p_G})= \pi_{0}(C_{p_G+n}).
\end{align}
The classifying spaces indeed satisfy the above highly non-trivial
relation.  This is the Bott periodicity theorem.  The theorem is obtained by
the following observation: the space $C_{p+1}$ can be viewed (in a homotopic
sense) as $\Om C_p$ -- the space of loops in $C_p$.  So we have
$\pi_1(C_p)=\pi_0(\Om C_p) = \pi_0(C_{p+1})$.  Similarly, the space $R_{p+1}$
can be viewed (in a homotopic sense) as $\Om R_p$ -- the space of loops in
$R_p$.  So we have $\pi_1(R_p)=\pi_0(\Om R_p) = \pi_0(R_{p+1})$.
 As a result
of Bott periodicity theorem, the classification of defects is independent of
spatial dimensions.  It only depends on the dimension and the symmetry of the
defects. If the defects lower the symmetry, then, we should use the reduced
symmetry to classify the defects.
In Tables \ref{defC} and \ref{defR}, we
list the classifications of those symmetric point and line defects for gapped
free fermion systems with various symmetries.  We would like to point out that
the line defects classified by $\Z$ in superconductors without symmetry
\emph{do not} correspond to the vortex lines (which usually belong to the
trivial class under our classification). The non-trivial line defect here
should carry chiral modes that only move in one direction along the defect
line.  In general, non-trivial defects have protected gapless excitations in
them.

\section{Summery}

In this paper, we study different possible full symmetry groups $G_f$ of
fermion systems that contain $U(1)$, time reversal $T$, and/or charge
conjugation $C$ symmetry.  We show that each symmetry group $G_f$ is associated
with a classifying space $C_{p_G}$ or $R_{p_G}$ (see Tables \ref{GFF} and
\ref{GFFsym}).  We classify $d$-dimensional gapped phases of free fermion
systems that have those full symmetry groups. We find that the different gapped
phases are described by $\pi_0(C_{p_G-d})$ or $\pi_0(R_{p_G-d})$.  Those
results, obtained using the K-theory approach, generalize the results in
\Ref{K0886,SRF0825} 

I would like thank A. Kitaev and Ying Ran for very helpful discussions.
This research is supported by NSF Grant No. DMR-1005541.


%

\end{document}